\documentclass{SCIS2021}

\begin{document}
\ArticleType{RESEARCH PAPER}
\Year{2020}
\Month{}
\Vol{}
\No{}
\DOI{}
\ArtNo{}
\ReceiveDate{}
\ReviseDate{}
\AcceptDate{}
\OnlineDate{}

\title{Uplink Transmission Design for Crowded Correlated Cell-Free Massive MIMO-OFDM Systems}{Uplink Transmission Design for Crowded Correlated Cell-Free Massive MIMO-OFDM Systems}

\author[1]{Junyuan GAO}{}
\author[1]{Yongpeng WU}{{yongpeng.wu@sjtu.edu.cn (Yongpeng Wu)}}
\author[2]{Yongjian WANG}{{wyj@cert.org.cn (Yongjian Wang)}}
\author[1]{Wenjun ZHANG}{}
\author[1]{Fan WEI}{}

\AuthorMark{Junyuan GAO}
\AuthorCitation{J. Gao, Y. Wu, Y. Wang, et al}

\address[1]{The Department of Electronic Engineering, Shanghai Jiao Tong University, Shanghai {\rm 200240}, China}
\address[2]{The National Computer Network Emergency Response Technical Team/Coordination Center of China,\\ Beijing {\rm 100029}, China}

\abstract{
  In cell-free massive multiple-input multiple-output (MIMO) orthogonal frequency division multiplexing (OFDM) systems, user equipments (UEs) are served by many distributed access points (APs), where channels are correlated due to finite angle-delay spread in realistic outdoor wireless propagation environments.
  Meanwhile, the number of UEs is growing rapidly for future fully networked society.
  In this paper, we focus on the uplink transmission design in crowded correlated cell-free massive MIMO-OFDM systems with limited number of orthogonal pilots.
  For the pilot transmission phase, we identify active UEs based on non-orthogonal pilot phase shift hopping patterns and non-orthogonal adjustable phase shift pilots (APSP).
  We derive a closed-form expression of mean square error of channel estimation (MSE-CE) and obtain an optimal condition for minimizing MSE-CE. According to this condition, the APSP set allocation scheme is proposed. Furthermore, for the data transmission, the max-min power control algorithm is devised to maximize the minimum spectral efficiency (SE) lower bound among active UEs.
  Simulation results indicate significant performance gains in terms of MSE-CE for the proposed APSP set allocation scheme. The proposed power control scheme can further improve the minimum SE among active UEs.
  Hence, they are crucial for crowded correlated cell-free massive MIMO-OFDM systems. }


\maketitle

\section{Introduction}
% 介绍分布的 cf massive MIMO和集中的 massive MIMO 的区别
  Massive multiple-input multiple-output (MIMO) has attracted great research interest in the past few decades and has been a fundamental structure in communication systems.
  Based on the antenna array deployment, massive MIMO architectures can be divided into co-located and distributed architectures.
  In co-located massive MIMO systems, the only base station (BS) serves all user equipments (UEs) in a cell~\cite{b2}.
  It is recognized as a key component for the fifth-generation (5G) wireless communication networks, which can increase spectral efficiency (SE) and energy efficiency (EE) with simple signal processing~\cite{b1}.
  However, these benefits are primarily enjoyed by cell-center UEs, but the performance of cell-edge UEs is usually limited by inter-cell interference.
  For distributed massive MIMO systems, access points (APs) equipped with single or multiple antennas are spread out over a large area~\cite{b3}.
  Cell-free massive MIMO is a special implementation of distributed massive MIMO, where many APs cooperate to jointly serve UEs and UEs experience no inter-cell interference during data downlink transmission, and hence the performance of cell-edge UEs can be greatly improved~\cite{zhang_jy}.
  Meanwhile, APs are connected to the central processing unit (CPU) via a fronthaul network.
  Thus, compared with co-located systems, cell-free massive MIMO needs more deployment costs and fronthaul overhead.
  However, cell-free massive MIMO systems have higher coverage probability since they can provide macro-diversity due to the joint coherent transmission from APs in a distributed topology, which is the key motivation for the study of cell-free massive MIMO~\cite{b4}.
  In order to further improve the system performance, cell-free massive MIMO is expected to be combined with orthogonal frequency division multiplexing (OFDM), which is a multi-carrier modulation technology with high data rate and high robustness to channel frequency selectivity~\cite{b5}.

% 介绍mmtc等多用户场景
  As we transition into the fully networked society, many use cases are emerging, like Internet of Things, social networking, and video streaming~\cite{wu,b6}. The massive machine-type communication (mMTC) has been one of the most representative services for future wireless communication systems. It aims to support massive connectivity of UEs which transmit packets in an intermittent pattern~\cite{b6}. Since the cell-free massive MIMO-OFDM can offer spatial degrees of freedom and macro-diversity, it can be utilized in crowded scenarios to improve the connectivity reliability of UEs.

  Channel state information (CSI) is crucial in wireless communication systems. Crowded scenarios pose new challenges in the pilot-based acquisition of CSI due to two reasons. First, the number of orthogonal pilots is limited by power budget and coherence interval. It is difficult to assign a dedicated pilot to each UE for channel estimation~\cite{b8,wuyp}. Second, UEs are sporadically active~\cite{b9}. In this case, pilot contamination becomes the performance bottleneck, and how to efficiently reduce channel estimation error and identify active UEs becomes an important topic.

\subsection{Motivation and Related Works}
  Nowadays, many works focus on the study of cell-free massive MIMO.
  For example, it was verified that cell-free massive MIMO can take advantage of the basic properties of co-located massive MIMO, including channel hardening and favorable propagation, when the number of antennas per AP was large and UEs were spatially well separated in independent and identically distributed (i.i.d.) fading channels~\cite{b10}.
  In~\cite{b4,b11}, the uplink and downlink SE of cell-free massive MIMO were derived with i.i.d. small-scale fading channels. In this case, cell-free massive MIMO can significantly outperform the small-cell operation in terms of the 95\%-likely per-user throughput~\cite{b11}, and can outperform the conventional single-cell massive MIMO in terms of the SE of cell-edge UEs~\cite{cellular,zhang_jy}.
  The uplink SE of different cell-free implementations are analyzed with spatially correlated fading channels~\cite{correlated_SE}.
  Taking the fronthaul power consumption into account, the EE was optimized via power control in i.i.d. fading channels~\cite{b12}.

  Some papers also study the pilot assignment in cell-free massive MIMO systems, but the vast majority of them consider i.i.d. fading channels and orthogonal pilot set.
  The most straightforward scheme is random pilot assignment (RPA), where each UE is randomly assigned a pilot from the orthogonal pilot set. It can lead to serious pilot contamination due to less consideration of the minimum distance among copilot UEs~\cite{b13}.
  In~\cite{b4}, a greedy pilot assignment scheme was utilized, where pilots were randomly allocated to UEs and then iteratively updated to improve the lowest rate among UEs.
  A structured pilot assignment (SPA) scheme was proposed in~\cite{b15} to maximize the minimum distance among copilot UEs.
  A tabu-based scheme was utilized to assign each UE a pilot from an orthogonal pilot set~\cite{tabu}.
  An efficient pilot assignment scheme based on graph coloring was proposed to mitigate the severe pilot contamination~\cite{graph_color}.

  To mitigate severe pilot contamination, most of existed works consider how to assign a pilot for each UE from the orthogonal pilot set in i.i.d. fading channels.
  However, in crowded scenarios where UEs are active in an intermittent pattern, it is unnecessary and inefficient to allocate each UE with a pilot from the orthogonal pilot set.
  Besides, in realistic outdoor wireless propagation environments, the channel angle-delay spread is limited due to the finite number of scattering clusters, i.e., practical channels between antennas of an AP and a UE over all subcarriers are spatially and frequently correlated~\cite{b9}.
  Hence, how to design uplink transmission scheme to identify active UEs, reduce channel estimation error, and efficiently transmit data in crowded cell-free massive MIMO-OFDM systems with  spatially and frequently correlated channels becomes an important issue.

\subsection{Contributions and Organization}
% 本文主要工作：随机接入 + cell-free Massive MIMO-OFDM + 相关信道
% 1. 考虑了三种相关性：空间相关性来源于扩散角度的有限性和不同的大尺度衰落，频率相关性来源于扩散时延的有限性；
%    建立了完善的信道模型，cell-free Massive MIMO、OFDM、相关信道、空间频率角度时延
% 2. 在 随机接入 + cell-free Massive MIMO-OFDM + 相关信道 场景下，推导出考虑了冲突模式、设备活跃模式、用户干扰、信道相关性、非正交导频集等因素的信道估计误差、频带利用效率和能量利用效率表达式；
% 3. 以使得信道估计误差最小的条件为基础，提出了基于导频集分配的随机接入算法，并分析了在不同角度扩展、时延扩展、AP 数目、AP选择比例场景下的系统性能；
% 4. 提出了上行链路用户的功率分配算法，与用户对AP的选择相结合，进一步提升系统的频带利用效率和能量利用效率。
Motivated by the aforementioned discussion, we consider the uplink pilot and data transmission in crowded cell-free massive MIMO-OFDM systems with spatially and frequently correlated channels. The main contributions of this paper are as follows:
\begin{itemize}
  \item
  We establish a comprehensive channel model with spatial and frequency correlation between antennas of an AP and a UE over all subcarriers resulting from finite angle-delay spread. We formulate the relationship between the space-frequency domain channel covariance matrix and the angle-delay domain channel power spectrum for cell-free massive MIMO-OFDM systems.
  \item
  We achieve the identification of active UEs based on non-orthogonal pilot sequences and non-orthogonal pilot phase shift hopping patterns.
  \item
  We derive the expressions of mean square error (MSE) of channel estimation (MSE-CE) and obtain the condition of minimizing MSE-CE. Based on this condition, we propose the adjustable phase shift pilot (APSP) set allocation scheme. Simulation results indicate its performance gain over other schemes.
  \item
  We devise a max-min power control algorithm to maximize the minimum SE lower bound of active UEs. We obtain the globally optimal solutions by iteratively solving linear programs.
\end{itemize}

  The rest of paper is organized as follows. Section \ref{section2} describes the system model for cell-free massive MIMO-OFDM systems. Section \ref{section3} presents the channel estimation, the identification of active UEs, and the APSP set allocation scheme.
  The expressions of SE and a lower bound of SE are derived in Section \ref{section4}. Here, the power control scheme is developed. Simulation results and discussions are given in Section \ref{section5}. Section \ref{section6} concludes the work.

  \emph{Notation:} We adopt uppercase and lowercase boldface letters to denote matrices and column vectors, respectively. We use $\left[\mathbf{A} \right]_{m,n}$, $\left[\mathbf{A} \right]_{m,:}$, and $\left[\mathbf{A} \right]_{:,n}$ to denote the $\left( m,n \right)$-th element, the $m$-th row vector, and the $n$-th column vector of matrix $\mathbf{A}$, respectively.
  We adopt $\mathbf{I}_{N\times N}$ to denote the $N\times N$ dimensional identity matrix, and $\mathbf{I}_{N\times G}$ to denote the matrix comprising the first $G \left( \leq N \right)$ columns of $\mathbf{I}_{N\times N}$.
  We use $\textbf{1}_{N\times 1}$ to denote all-one vector, and $\textbf{0}$ to denote all-zero vector or matrix.
  $\left\|\cdot \right\|$ denotes the Euclidean norm.
  The conjugate, transpose, and conjugate transpose are denoted by $\left(\cdot \right)^{*}$, $\left(\cdot \right)^{T}$, and $\left(\cdot \right)^{H}$, respectively.
  $\mathbb{E}\{\cdot \}$ denotes the expectation operator.
  The notation $\rm{vec}$, $\otimes$, and $\odot$ denote vectorization, Kronecker product, and Hadamard product, respectively.
  $\backslash$ denotes the set subtraction.
  $\left| \mathcal{A} \right|$ denotes the cardinal of set $\mathcal{A}$.
  Modulo-\emph{N} is denoted by $\left\langle \cdot \right\rangle_{N}$.
  $\lfloor x\rfloor$ denotes the largest integer not greater than $x$.
  $z\sim\mathcal{CN}(0,\sigma^{2})$ denotes a circularly symmetric complex Gaussian random variable (RV) with mean $0$ and variance $\sigma^{2}$.
  Let $\delta(x) =\left\{\begin{array}{ll}
  1, & x=0 \\
  0, & x\neq 0
  \end{array}\right.$.
  Let $\operatorname{diag}\left\{\mathbf{a}\right\}$ denote a diagonal matrix with vector $\mathbf{a}$ comprising its diagonal elements. $\operatorname{diag}\left\{ \mathbf{A},\mathbf{B} \right\}$ denotes a block diagonal matrix with $\mathbf{A}$ and $\mathbf{B}$ in diagonal blocks.

\section{System Model} \label{section2}
\begin{figure}[!t]
\centering
\includegraphics[width=0.45\linewidth]{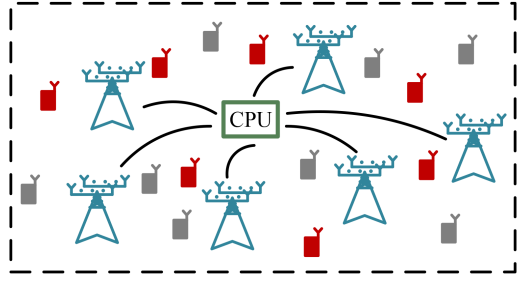}
\caption{Cell-free massive MIMO system. Active and inactive UEs are in red and gray, respectively.}
\label{fig:1}
\end{figure}

  We consider an uplink cell-free massive MIMO-OFDM system where \emph{L} APs serve a maximal number of \emph{K} single-antenna UEs in the same time-frequency resource under the time-division duplex (TDD) mode.
  APs and UEs are randomly located over a large area.
  We assume each AP is equipped with two uniform linear arrays (ULAs).
  Each ULA comprises of \emph{N} directional antennas and receives waves with the incidence angle in a range of $180$ degrees to avoid ambiguities in the array manifold.
  We denote  $M\!=\!NL$.
  The sets of APs and UEs are denoted by  $\mathcal{L}\!=\!\{0,1,\cdots\!,L\!-\!1\}$ and $\mathcal{K}\!=\!\{0,1,\cdots\!,K\!-\!1\}$, respectively. The set of active UEs is denoted as $\mathcal{K}_{a}$ with the size of $K_a\leq K$.
  APs are connected to a CPU via an error-free fronthaul network, as shown in Fig.~\ref{fig:1}.

  We adopt OFDM with $N_c$ subcarriers to combat time dispersive channels. The sampling duration is $T_s$. We employ $T_{\rm{sym}}=\left(N_{c}\!+\!N_{\rm{cp}}\right)T_{s}$ and $T_{c}\!=\!N_{c}T_{s}$ to denote the OFDM symbol duration with and without cyclic prefix (CP), where $N_{\rm{cp}}$ denotes the number of symbols in CP. The CP duration $T_{\rm{cp}}\!=\!N_{\rm{cp}}T_{s}$ is assumed to be longer than the maximum channel delay of UEs.

\subsection{Channel Model in Space-Frequency Domain}
  In realistic outdoor wireless propagation environments, the channel angle spread is limited due to finite scattering clusters. Hence, we assume two ULAs in an AP separately receive waves from UEs with the mean angle of arrival (AoA)
  in a range of $180$ degrees,
  i.e., the waves from a UE are received by directional antennas of a ULA in each AP.
  Let $\mathbf{g}_{k, l, s}^{\beta}$ denote the uplink channel response vector between the \emph{l}-th AP and the \emph{k}-th UE over the \emph{s}-th subcarrier.
  We employ $\theta$ to denote the incidence angle. $\mathbf{g}_{k, l, s}^{\beta} $ can be modeled as~\cite{b17,b18,b19}
    \begin{equation}\label{g_kls}
      \mathbf{g}_{k, l, s}^{\beta} \triangleq \sqrt{\beta_{k, l}}\; \mathbf{g}_{k, l, s} = \sum_{q=0}^{N_{\rm{cp}}-1} \int_{-\frac{\pi}{2}}^{\frac{\pi}{2}}  \mathbf{v}_{N}(\theta) \exp \left\{-\overline{\jmath} 2 \pi \frac{s}{T_{c}} \tau\right\} a_{k, l}^{\beta}(\theta, \tau) \delta \left(\tau-q T_{s}\right) \mathrm{d} \theta \in \mathbb{C}^{N\times1},
    \end{equation}
  where $\beta_{k, l}$ denotes the large-scale fading between the \emph{l}-th AP and the \emph{k}-th UE, $\mathbf{g}_{k, l, s}$ denotes the small-scale fading, and $a_{k, l}^{\beta}(\theta, \tau) \triangleq  \sqrt{\beta_{k, l}} a_{k, l}(\theta, \tau)$ represents the angle-delay domain channel gain function corresponding to the incidence angle $\theta$ and delay $\tau$. If AP antennas are spaced with half of wavelength, the steering vector $\mathbf{v}_N(\theta)=[1,e^{-\overline{\jmath}\pi\sin\theta} ,\cdots ,e^{-\overline{\jmath}\pi(N-1)\sin\theta}]^T$.
  The space-frequency domain channel response matrix between UE \emph{k} and all APs over $N_c$ subcarriers is denoted as
    \begin{equation}\label{G_k}
      \mathbf{G}_{k}^{\beta} \triangleq
      \begin{bmatrix} {\mathbf{G}_{k, 0}^{\beta}}\\ {\mathbf{G}_{k, 1}^{\beta}} \\{\vdots} \\{\mathbf{G}_{k, L -1}^{\beta}} \end{bmatrix} =
      \begin{bmatrix}
      {\mathbf{g}_{k, 0, 0}^{\beta}}  & {\mathbf{g}_{k, 0, 1}^{\beta}}  &  \cdots  &  {\mathbf{g}_{k, 0, N_c-1}^{\beta}} \\
      {\mathbf{g}_{k, 1, 0}^{\beta}}  & {\mathbf{g}_{k, 1, 1}^{\beta}}  &  \cdots  &  {\mathbf{g}_{k, 1, N_c-1}^{\beta}} \\
      {\vdots}  & {\vdots}  & {\ddots}  &  {\vdots} \\
      {\mathbf{g}_{k, L-1, 0}^{\beta}}  & {\mathbf{g}_{k, L-1, 1}^{\beta}}  &  \cdots  &  {\mathbf{g}_{k, L-1, N_c-1}^{\beta}}
      \end{bmatrix}
      = \begin{bmatrix} {\mathbf{g}_{k, 0}^{\beta}}& {\mathbf{g}_{k, 1}^{\beta}}& {\cdots} & {\mathbf{g}_{k, N_c-1}^{\beta}} \end{bmatrix} \in  \mathbb{C}^{M \times N_{c}},
    \end{equation}
  where ${\mathbf{G}_{k, l}^{\beta}}\in \mathbb{C}^{N \times N_{c}}$ and ${\mathbf{g}_{k, s}^{\beta}}\in \mathbb{C}^{M \times 1}$ are the space-frequency domain channel response matrix between the \emph{l}-th AP and the \emph{k}-th UE over $N_c$ subcarriers, and the vector between all APs and the \emph{k}-th UE over the \emph{s}-th subcarrier, respectively.
  It is assumed that $\operatorname{vec}\left\{\mathbf{G}_{k}^{\beta}\right\}\sim \mathcal{CN}\left(0,\mathbf{R}_k^{\beta}\right)$.

  In this work, we consider the space-frequency domain correlation between the antennas of an AP and a UE over $N_c$ subcarriers.
  We assume different UEs or APs are spatially separated by a few wavelengths, and the channel realizations between different UEs and an AP and the channel realizations between a UE and different APs are uncorrelated~\cite{Adhikary}.
  Besides, we assume the angle-delay domain channel gain function ${a_{k,l}(\theta,\tau)}$ is uncorrelated~\cite{b10,b18}, i.e.,
    \begin{equation}
      \mathbb{E}\left\{a_{k, l}(\theta, \tau) a^{\ast}_{k', l^{\prime}}\left(\theta^{\prime}, \tau^{\prime}\right) \right\} = P^{A}_{k, l}(\theta) P^{D}_{k, l}(\tau)
      \delta\left(k-k^{\prime}\right)\delta\left(l-l^{\prime}\right) \delta\left(\theta-\theta^{\prime}\right) \delta\left(\tau-\tau^{\prime}\right),
    \end{equation}
  where $P^{A}_{k, l}(\theta)$ and $P^{D}_{k, l}(\tau)$ represent the channel power azimuth spectrum (PAS) and power delay spectrum (PDS), respectively.
  Let $\varsigma_{k,l}$, $\theta_{k,l}$, and $\zeta_{k,l}$ denote the angle spread, mean AoA, and delay spread between the \emph{k}-th UE and the \emph{l}-th AP, respectively.
  PAS and PDS are given by~\cite{b20}
    \begin{equation}\label{PAS}
      P^{A}_{k, l}(\theta) \propto \exp \left\{ -{\sqrt{2}\left|\theta-\theta_{k, l}\right|} / {\varsigma_{k, l}}\right\}, \text { for } \theta, \theta_{k, l} \in [-\pi/2,\pi/2],
    \end{equation}
    \begin{equation}\label{PDS}
      P^{D}_{k, l}(\tau) \propto \exp \left\{-{\tau}/ {\zeta_{k,l}} \right\}, \text { for } \tau \in\left[0, \left(N_{\mathrm{cp}}-1\right)T_s \right].
    \end{equation}
  We assume the channel angle spreads between different UEs and APs are equal, i.e., $\varsigma_{k,l}=\varsigma$ for $k\in \mathcal{K}$ and $l \in \mathcal{L}$. Similarly, the delay spreads satisfy $\zeta_{k,l}=\zeta$ for $k\in \mathcal{K}$ and $l \in \mathcal{L}$.

\subsection{Channel Model in Angle-Delay Domain}
  Let $\mathbf{H}^{\beta}_{k,l}\triangleq\sqrt{\beta_{k,l}}\mathbf{H}_{k,l} \in \mathbb{C}^{N \times N_{\mathrm{cp}}}$ denote the angle-delay domain channel response matrix between the \emph{k}-th UE and the \emph{l}-th AP.
  Let $ \mathbf{H}^{\beta}_{k} \triangleq \left[  \left(\mathbf{H}^{\beta}_{k,0}\right)^{T} \left(\mathbf{H}^{\beta}_{k,1}\right)^{T}\cdots \left(\mathbf{H}^{\beta}_{k,L-1}\right)^{T} \right]^{T}$ and $\mathbf{H}_{k} \triangleq \left[ \mathbf{H}_{k,0}^{T}  \;\mathbf{H}_{k,1}^{T}\cdots \mathbf{H}_{k,L-1}^{T} \right]^{T}$ denote the angle-delay domain channel response matrices between UE \emph{k} and all APs with and without large-scale fading, respectively.
  Different elements in $\mathbf{H}^{\beta}_{k}$ correspond to the channels for different incidence angles, delays or APs, which satisfy~\cite{b10,b18}
    \begin{equation}\label{angle-delay power}
      \mathbb{E}\left\{\left[\mathbf{H}^{\beta}_{k}\right]_{m, q}\left[\mathbf{H}^{\beta}_{k}\right]_{m^{\prime}, q^{\prime}}^{*}\right\}=\delta\left(m-m^{\prime}\right) \delta\left(q-q^{\prime}\right) \left[\boldsymbol{\Upsilon}^{\beta}_{k}\right]_{m, q},
    \end{equation}
  where $\boldsymbol{\Upsilon}_{k}^{\beta} \!\triangleq\!
      \begin{bmatrix}
        \left(\boldsymbol{\Upsilon}^{\beta}_{k,0}\right)^{\!T} \!\!\!& \left(\boldsymbol{\Upsilon}^{\beta}_{k,1}\right)^{\!T} \!\!\!&\! \cdots \!\!\!& \left(\boldsymbol{\Upsilon}^{\beta}_{k,L-1}\right)^{\!T}
      \end{bmatrix}^{\!T} \!\in\! \mathbb{R}^{M\times N_{\mathrm{cp}}}$ is the angle-delay domain channel power spectrum between UE \emph{k} and all APs. We define $\boldsymbol{\Upsilon}_{k} \!\triangleq\!
      \begin{bmatrix}
         \boldsymbol{\Upsilon}^{T}_{k,0} \!& \boldsymbol{\Upsilon}^{T}_{k,1} \!\!&\! \cdots \!\!&\! \boldsymbol{\Upsilon}^{T}_{k,L-1}
      \end{bmatrix}^{\!T} \!\in\! \mathbb{R}^{M\times N_{\mathrm{cp}}}$.

    Wireless channels are sparse in many typical scenarios~\cite{b21,b22}. In this work, we consider the angle-delay domain sparsity from a UE to all APs in a cell-free massive MIMO-OFDM system. Most elements in $\boldsymbol{\Upsilon}_{k}^{\beta}$ are approximately 0 due to finite angle-delay spread.

\subsection{Relationship between Space-Frequency Domain and Angle-Delay Domain Channels}
  Since the channel power lies in a finite number of delays due to limited scattering, channels are frequently correlated~\cite{b18}. Considering most channel power is concentrated in a finite region of angles, channels between the antennas of an AP and a UE are spatially correlated~\cite{b22}. Proposition~\ref{prop_1} shows the relationship between the space-frequency domain channel covariance matrix and the angle-delay domain channel power spectrum in cell-free massive MIMO-OFDM systems.
  \proposition[] \label{prop_1}
    {Define $\theta_n \!=\! \arcsin\left(\frac{2n}{N}-1\right)$ for $n=0,1,\cdots\!,N$.
    Let $\left[\mathbf{V}_{N\times N}\right]_{i, j} \triangleq \frac{1}{\sqrt{N}} \exp \left\{-\overline{\jmath} \pi i \sin\theta_n  \right\}$,
    $\mathbf{V}_{  M\times M} \triangleq  \mathbf{I}_{L\times L} \otimes \mathbf{V}_{  N\times N}$, and
    $\left[\mathbf{F}_{N_{c} \times N_{\mathrm{cp}}}\right]_{s,q}  \triangleq \frac{1}{ \sqrt{N_{c}}} \exp  \left\{-\overline{\jmath} 2 \pi \frac{s}{N_{c}}q\right\}$.
    We have $\mathbf{V}_{N\times N}^{H}\mathbf{V}_{N\times N} \stackrel{N \rightarrow \infty}{=} \mathbf{I}_{N\times N}$ and $\mathbf{V}_{M\times M}^{H}\mathbf{V}_{M\times M} \stackrel{N \rightarrow \infty}{=} \mathbf{I}_{M\times M}$.
    The channel power spectrum between the \emph{k}-th UE and the \emph{l}-th AP is denoted as $\boldsymbol{\Upsilon}_{k,l}^{\beta} \in \mathbb{R}^{N\times N_{\mathrm{cp}}}$ satisfying
      \begin{equation}
        \left[\boldsymbol{\Upsilon}_{k,l}^{\beta}\right]_{n, q} \triangleq \beta_{k,l}\left[\boldsymbol{\Upsilon}_{k,l}\right]_{n, q} = \beta_{k,l} N N_{c}\left(\theta_{n+1}-\theta_{n}\right) P^{A}_{k, l}(\theta_{n}) P^{D}_{k, l}(\tau_{q}),
      \end{equation}
    where $\tau_{q}=q T_{s}$ for $q=0,1,\cdots,N_{\mathrm{cp}}-1$.
    The relationship between the space-frequency domain channel covariance matrix and angle-delay domain channel power spectrum is given by
      \begin{equation} \label{relationship_R}
        \mathbf{R}^{\beta}_k \stackrel{N \rightarrow \infty}{=} \left(\mathbf{F}_{N_{c} \times N_{\mathrm{cp}}} \otimes \mathbf{V}_{M\times M}\right) \operatorname{diag}\bigg\{\operatorname{vec} \left\{\boldsymbol{\Upsilon}_{k}^{\beta}\right\}\bigg\}\left(\mathbf{F}_{N_{c} \times N_{\mathrm{cp}}} \otimes \mathbf{V}_{M\times M}\right)^{H}  \in \mathbb{C}^{M  N_{c} \times M  N_{c}}.
      \end{equation}
    The space-frequency domain channel response matrix $\mathbf{G}_{k}^{\beta}$ can be decomposed as
      \begin{equation}\label{relationship_G}
        \mathbf{G}_{k}^{\beta} \stackrel{N \rightarrow \infty}{=} \mathbf{V}_{M\times M} \mathbf{H}_{k}^{\beta} \;\mathbf{F}^{T}_{N_{c} \times N_{\mathrm{cp}}}.
      \end{equation}
    }
    \proof {See \ref{proof_prop_1}.}

  When $N$ is sufficiently large, the eigenvectors of space-frequency domain channel covariance matrices are independent of the locations of UEs and only depend on the AP antenna configurations. Eigenvalues can be approximated by the angle-delay domain channel power spectrum.
  Specifically, when ULAs are employed at each AP with antenna spacing of half-wavelength, $\mathbf{V}_{N\times N}$ can be set to discrete Fourier transform (DFT) matrices with some matrix elementary operations.
  Since the dimension of $\boldsymbol{\Upsilon}_{k}^{\beta}$ is much smaller than that of $\mathbf{R}_k^{\beta}$ and most elements in $\boldsymbol{\Upsilon}_{k}^{\beta}$ are approximately 0, we will estimate angle-delay domain channel parameters at first, and then obtain space-frequency domain channel parameters via \eqref{relationship_G}. We assume angle-delay domain channel power spectrums are known by APs.

  Proposition~\ref{prop_1} is a generalization of existing results. It agrees with the results in~\cite{b22} where channels are frequency-flat fading on a narrow-band subcarrier.
  It is consistent with the results in~\cite{b19}, where correlated channels in co-located systems are considered.
  The approximation is shown to be accurate enough when the number of antennas at the BS ranges from 64 to 512 in frequency-flat co-located massive MIMO channels~\cite{b23}.
  %In this work, we extend the relationship between the space-frequency domain channels and angle-delay domain channels from existing co-located systems to cell-free massive MIMO-OFDM systems with spatial and frequency correlation.

\section{Pilot Transmission Design with Pilot Assignment} \label{section3}
  UEs are sporadically active and far more than orthogonal pilots in crowded scenarios, which calls for proper design of uplink transmission scheme and pilot assignment scheme.
  We utilize APSP to increase the number of unique pilots as shown in Section~\ref{section3_sub1}.
  Besides, active UEs can be identified through pilot shift hopping patterns as introduced in Section~\ref{section3_sub2}.
  To better utilize channel characteristics and improve MSE-CE in correlated cell-free massive MIMO-OFDM systems, an APSP set allocation scheme is proposed in Section~\ref{section3_sub3}.
\subsection{APSP and Channel Estimation}\label{section3_sub1}
  We assume channels are piece-wise constant over a coherence interval. The time coherence interval is assumed to be equal to the length of $Z_a$ OFDM symbols. The frequency coherence interval is assumed to be equal to $\frac{1}{2 \zeta }$. We assume $Z$ OFDM symbols are used for pilot transmission and the remaining $Z_a-Z$ OFDM symbols are for uplink data transmission.

  During the training phase, all active UEs simultaneously transmit pilots to APs. The space-frequency domain signal received at the \emph{l}-th AP is given by
  \begin{equation}\label{receive_Y}
    \mathbf{Y}_{l}=\sqrt{\rho_{p}Z} \sum_{k^{\prime} \in \mathcal{K}_{a}} \mathbf{G}_{k^{\prime}, l}^{\beta} \mathbf{\Phi}_{k^{\prime}}+\mathbf{W}_{l}
    \in \mathbb{C}^{N \times N_{c}Z},
  \end{equation}
  where $\rho_{p}$ is the normalized signal-to-noise ratio (SNR) in the training phase, $\mathbf{W}_{l}$ is the additive white Gaussian noise (AWGN) matrix including i.i.d. $\mathcal{CN} (0,1)$ elements, $\mathbf{G}_{k^{\prime}, l}^{\beta} = \mathbf{V}_{N\times N} \mathbf{H}_{k^{\prime},l}^{\beta} \mathbf{F}^{T}_{N_{c} \times N_{\mathrm{cp}}}$, and $\mathbf{\Phi}_{k^{\prime}}$ is the pilot signal given in~\eqref{pilot_k}. We assume large-scale fading coefficients are known wherever required.

  In cell-free massive MIMO-OFDM systems, the number of phase shift orthogonal pilots (PSOPs) is approximately $\lfloor Z N_c/N_{cp}\rfloor$. Their phase shift differences are no less than the maximum channel delay of UEs to promise orthogonality of different pilots~\cite{b24}. However, pilot contamination can be serious since UEs are far more than PSOPs. Hence, APSP is adopted in this work~\cite{b19}. The pilot signal for the \emph{k}-th UE over $N_c$ subcarriers and $Z$ OFDM symbols is
  \begin{equation}\label{pilot_k}
    \mathbf{\Phi}_{k} \triangleq \left[ \mathbf{U}\right]_{\left\langle \phi_{k} \right\rangle_{Z},:} \otimes
    \left( \mathbf{D}_{\left\lfloor \phi_{k}/Z\right\rfloor} \mathbf{\Phi}_b \right) \in \mathbb{C}^{N_{c} \times N_{c}Z},
  \end{equation}
  where $\mathbf{U}$ is an arbitrary $Z\times Z$ dimensional unitary matrix,
  $\phi_{k} \in \mathbf{\Psi} = \left\{ 0,1,\cdots,N_cZ-1\right\}$,
  $\mathbf{D}_{i} = \operatorname{diag}\left\{\mathbf{f}_{N_{c}, i}\right\} $,
  $\mathbf{f}_{N_{c}, i} = \left[1 \; \exp\left\{-\overline{\jmath} 2 \pi \frac{i}{N_{c}} \right\}  \cdots  \exp\left\{-\overline{\jmath} 2 \pi \frac{\left(N_{c}-1 \right)i}{N_{c}} \right\} \right]^{T}$,
  and $\mathbf{\Phi}_b$ is a diagonal matrix satisfying $\mathbf{\Phi}_b \mathbf{\Phi}_b^H = \mathbf{I}_{N_{c} \times N_{c}}$, which is the basic pilot matrix shared by all UEs.
  Then we have, for $\forall k, k^{\prime} \in \mathcal{K}$,
  \begin{equation} \label{pilot_k_k1}
    \mathbf{\Phi}_{k'} \mathbf{\Phi}_{k}^{H}
    \stackrel{(\mathrm{a})}{=} \left([\mathbf{U}]_{\left\langle \phi_{k'} \right\rangle_{Z},:}[\mathbf{U}]_{\left\langle \phi_{k} \right\rangle_{Z},:}^{H}\right) \otimes\left(\mathbf{D}_{\left\lfloor \phi_{k'}/Z\right\rfloor} \mathbf{D}_{\left\lfloor \phi_{k}/Z\right\rfloor}^{H}\right)
    =\delta\left(\left\langle\phi_{k^{\prime}}\right\rangle_{Z}-\left\langle\phi_{k}\right\rangle_{Z}\right) \mathbf{D}_{\left\lfloor{\phi_{k'}}/{Z}\right\rfloor - \left\lfloor{\phi_{k}}/{Z}\right\rfloor},
  \end{equation}
  where (a) follows from the properties of Kronecker product, i.e., $(\mathbf{A} \otimes \mathbf{B})(\mathbf{C} \otimes \mathbf{D})=(\mathbf{A} \mathbf{C}) \otimes(\mathbf{B D})$ and $(\mathbf{A} \otimes \mathbf{B})^{H}=\mathbf{A}^{H} \otimes \mathbf{B}^{H}$.
  For the APSP, phase shifts are divided into $Z$ groups. The group index is $\left\langle\phi_{k}\right\rangle_{Z}$. There is no pilot interference if UEs use phase shifts in different groups.

  Based on \eqref{receive_Y}, \eqref{pilot_k}, and \eqref{pilot_k_k1}, after decorrelation, we have
  \begin{align}\label{receive_Y2}
    \check{\mathbf{Y}}_{k, l}
    &=\frac{1}{\sqrt{\rho_{p} Z\beta_{k, l}}} \mathbf{V}_{N
    \times N}^{H} \mathbf{Y}_{l} \mathbf{\Phi}_{k}^{H} \mathbf{F}_{N_{c} \times N_{\mathrm{cp}}}^{\ast} \notag\\
    &={{\mathbf{H}}_{k,l}}
    +\frac{1}{\sqrt{{{\beta }_{k,l}}}}\sum\limits_{{k}^{\prime}\in {{\mathcal{K}}_{a}}\backslash \left\{ k \right\}}
    \!\!\!\!\delta\left(\left\langle\phi_{k^{\prime}}\right\rangle_{Z}-\left\langle\phi_{k}\right\rangle_{Z}\right)
    \mathbf{H}_{{k}^{\prime},l}^{\beta }\mathbf{F}_{{{N}_{c}}\times {{N}_{\mathrm{cp}}}}^{T}
    \mathbf{D}_{\left\lfloor{\phi_{k'}}\!/{Z}\right\rfloor - \left\lfloor{\phi_{k}}/{Z}\right\rfloor}
    \mathbf{F}_{\!{{N}_{c}}\times {{N}_{\mathrm{cp}}}}^{\ast}
    +\frac{1}{\sqrt{{{\rho }_{p}}Z{{\beta }_{k,l}}}}{\check{\mathbf{W}}_{k,l}} \notag\\
    &={{\mathbf{H}}_{k,l}}
    +\frac{1}{\sqrt{{{\beta }_{k,l}}}}\sum\limits_{{k}^{\prime}\in {{\mathcal{K}}_{a}}\backslash \left\{ k \right\}}
    \!\!\!\!\delta\left(\left\langle\phi_{k^{\prime}}\right\rangle_{Z}-\left\langle\phi_{k}\right\rangle_{Z}\right)
    \mathbf{H}_{{k}^{\prime},l}^{\beta , \left\lfloor{\phi_{k'}}/{Z}\right\rfloor - \left\lfloor{\phi_{k}}/{Z}\right\rfloor }
    +\frac{1}{\sqrt{{{\rho }_{p}}Z{{\beta }_{k,l}}}}{\check{\mathbf{W}}_{k,l}},
  \end{align}
  where ${\check{\mathbf{W}}_{k,l}}$ includes i.i.d. $\mathcal{CN} (0,1)$ entrys due to unitary transformation.
  Define the permutation matrix $\mathbf{\Gamma}^{i}_{N_{c}}
  =\left[\begin{array}{cc}{\mathbf{0}} & {\mathbf{I}_{\left(N_{c}-\left\langle i\right\rangle_{N_{c}}\right) \times  \left(N_{c}-\left\langle i\right\rangle_{N_{c}}\right)}} \\ {\mathbf{I}_{ \left\langle i\right\rangle_{N_{c}} \times \left\langle i\right\rangle_{N_{c}}  }} & { \mathbf{0}} \end{array} \right]$.
  Then, $\mathbf{H}_{{k}^{\prime},l}^{\beta , \left\lfloor{\phi_{k'}}/{Z}\right\rfloor - \left\lfloor{\phi_{k}}/{Z}\right\rfloor }$ can be expressed as
  \begin{equation}\label{h_shift}
    \mathbf{H}_{{k}^{\prime},l}^{\beta , \left\lfloor{\phi_{k'}}/{Z}\right\rfloor - \left\lfloor{\phi_{k}}/{Z}\right\rfloor }
    = \mathbf{H}_{k^{\prime},l}^{\beta} \mathbf{I}_{N_{c} \times N_{\mathrm{cp}}}^{T} \mathbf{\Gamma}^{{\left\lfloor{\phi_{k'}}/{Z}\right\rfloor - \left\lfloor{\phi_{k}}/{Z}\right\rfloor}}_{N_{c}} \mathbf{I}_{N_{c} \times N_{\mathrm{cp}}}.
  \end{equation}
  The power spectrum of \eqref{h_shift} is given by
  \begin{align}\label{p_shift}
    \mathbf{\Upsilon}_{{{k}^{\prime},l}}^{\beta ,{\left\lfloor {\phi_{k'}} /{Z} \right\rfloor - \left\lfloor{\phi_{k}}/{Z} \right\rfloor}}
    &=\mathbb{E}\left\{ \mathbf{H}_{{k}^{\prime},l}^{\beta ,{\left\lfloor{\phi_{k'}}/{Z} \right\rfloor- \left\lfloor{\phi_{k}}/{Z}\right\rfloor} }
    \odot {{\left( \mathbf{H}_{{k}^{\prime},l}^{\beta , {\left\lfloor{\phi_{k'}}/{Z}\right\rfloor - \left\lfloor{\phi_{k}}/{Z}\right\rfloor}} \right)}^{\ast}} \right\}\notag\\
    &=\mathbf{\Upsilon}_{k^{\prime},l}^{\beta} \mathbf{I}_{N_{c} \times N_{\mathrm{cp}}}^{T} \mathbf{\Gamma}_{{{N}_{c}}}^{{\left\lfloor{\phi_{k'}}/{Z}\right\rfloor - \left\lfloor{\phi_{k}}/{Z}\right\rfloor}}{{\mathbf{I}}_{{{N}_{c}}\times {{N}_{\mathrm{cp}}}}}.
  \end{align}
  Define ${\mathbf{\Upsilon}}_{k'}^{\beta ,{\left\lfloor {\phi_{k'}} /{Z} \right\rfloor - \left\lfloor{\phi_{k}}/{Z} \right\rfloor}}  \triangleq  \begin{bmatrix}
       \left( {\mathbf{\Upsilon}}_{k',0}^{\beta ,{\left\lfloor {\phi_{k'}} /{Z} \right\rfloor - \left\lfloor{\phi_{k}}/{Z} \right\rfloor} } \right)^{ T}
      & \left( {\mathbf{\Upsilon}}_{k',1}^{\beta ,{\left\lfloor {\phi_{k'}} /{Z} \right\rfloor - \left\lfloor{\phi_{k}}/{Z} \right\rfloor} } \right)^{ T}  &  \cdots  &  \left( {\mathbf{\Upsilon}}_{k',L-1}^{\beta ,{\left\lfloor {\phi_{k'}} /{Z} \right\rfloor - \left\lfloor{\phi_{k}}/{Z} \right\rfloor} } \right)^{ T}  \end{bmatrix}^{T}$.

  Channels are estimated in a decentralized fashion in cell-free massive MIMO-OFDM systems, i.e., each AP autonomously estimates channels without interchanging information~\cite{b4}. For an AP, the channel estimation is performed in an element-wise manner in the angle-delay domain. The minimum mean-square error (MMSE) estimate of $\left[{{\mathbf{H}}_{k,l}}\right]_{i,j}$ is
  \begin{equation} \label{H_est}
    {{\left[ {{\widehat{\mathbf{H}}}_{k,l}} \right]}_{i,j}}
    =\frac{{{\left[ \mathbf{\Upsilon }_{k,l}^{\beta } \right]}_{i,j}}}{\sum\limits_{{k}^{\prime}\in {{\mathcal{K}}_{a}}}\delta\left(\left\langle\phi_{k^{\prime}}\right\rangle_{Z}-\left\langle\phi_{k}\right\rangle_{Z}\right)
    {{{\left[ \mathbf{\Upsilon }_{{k}',l}^{\beta ,{\left\lfloor\!{\phi_{k'}}/{Z}\right\rfloor - \left\lfloor\!{\phi_{k}}/{Z}\right\rfloor} } \right]}_{i,j}}+\frac{1}{{{\rho }_{p}Z}}}}{{\left[ {\check{\mathbf{Y}}_{k,l}} \right]}_{i,j}}.
  \end{equation}

  Denote by ${{\widetilde{\mathbf{H}}}_{k,l}} = {{{\mathbf{H}}}_{k,l}} - {{\widehat{\mathbf{H}}}_{k,l}}$  the channel estimation error. Based on the property of MMSE estimation, ${{\widetilde{\mathbf{H}}}_{k,l}}$ is independent of ${{\widehat{\mathbf{H}}}_{k,l}}$, and the MSE-CE is shown as
  \begin{equation} \label{MSE_CE_unaveraged}
    {{\left[ {\mathbf{\Xi}}_{k,l} \right]}_{i,j}} \!\triangleq
    \mathbb{E}\left\{{{\left[ {{\widetilde{\mathbf{H}}}_{k,l}} \right]}_{i,j}} {{\left[ {{\widetilde{\mathbf{H}}}_{k,l}} \right]}^{\ast}_{i,j}}\right\}
    = \left[ {{{\mathbf{\Upsilon}}}_{k,l}}\right]_{i,j}-
    \frac{{{\left[ {{\mathbf{\Upsilon }}_{k,l}} \right]}_{i,j}}{{\left[ \mathbf{\Upsilon }_{k,l}^{\beta } \right]}_{i,j}}}{\sum\limits_{{k}^{\prime}\in {{\mathcal{K}}_{a}}}{\!\delta\left(\left\langle\phi_{k^{\prime}}\right\rangle_{\!Z}
    \!-\left\langle\phi_{k}\right\rangle_{Z}\right)
    {{\left[ \mathbf{\Upsilon }_{{k}^{\prime},l}^{\beta ,{\left\lfloor{\phi_{k'}}/{Z}\right\rfloor - \left\lfloor{\phi_{k}}/{Z}\right\rfloor} } \right]}_{\!i,j}}}\!+\!\frac{1}{{{\rho }_{p}}Z}}.
  \end{equation}
  Define $ \boldsymbol{\Xi}_{k}^{\beta}  \triangleq  \begin{bmatrix}
      \beta_{k,0}\boldsymbol{\Xi}_{k,0}^{T}  &   \beta_{k,1}\boldsymbol{\Xi}_{k,1}^{T}  & \cdots  &  \beta_{k,L-1}\boldsymbol{\Xi}_{k,L-1}^{T} \end{bmatrix}^{T} $.
  \begin{figure}
  \centering
  \includegraphics[width=0.75\linewidth]{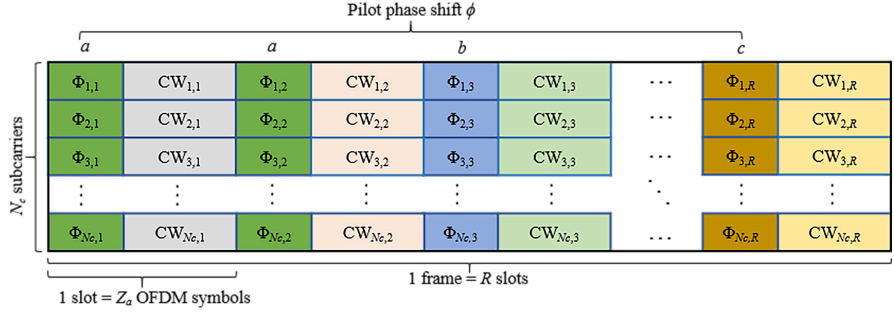}\\
  \caption{Illustration of the transmission frame. In this example, the hopping pattern for a given UE is $\left\{a,a,b,\cdots,c\right\}$. Pilot signals among $N_c$ subcarriers are formulated based on this pattern, and transmitted during the pilot transmission phase. Data codewords are transmitted afterwards.}
  \label{fig:2}
  \end{figure}

\subsection{UE Identification}\label{section3_sub2}
  Pilot contamination is unavoidable in crowded scenarios. In this case, transmitting UEs cannot be identified in a slot, and we have to resort to the information among some slots.
  In this work, the uplink transmission phase is divided into frames consisting of $R$ slots, as shown in Fig.~\ref{fig:2}. Each slot comprises $Z_a$ OFDM symbols.
  The phase shifts of pilot signals among $R$ slots, i.e., the hopping patterns, form the identification basis of active UEs.
  When PSOPs are utilized, the number of UEs that can be identified is $\lfloor Z N_c/N_{cp}\rfloor^{R}$. It can be further increased if APSPs are adopted due to channel sparsity. Hence, we assume each UE has a unique and non-orthogonal pseudo-random hopping pattern, which are known by APs and CPU in advance.

  Assuming the activation probability of UEs is $p_a$, the probability of having $K_a$ active UEs among \emph{K} UEs is given by
    \begin{equation}\label{K_a}
      p(K_a|K) = {K \choose {K_a}}p_a^{K_a}(1-p_a)^{K-K_a}.
    \end{equation}
  When a UE is active, it selects the pilot phase shift based on its hopping pattern. Next, it formulates corresponding pilot over $N_c$ subcarriers based on (\ref{pilot_k}) and transmits this pilot during the pilot transmission phase.
  Data codewords are transmitted afterwards. At the receiver, APs identify hopping patterns by running a correlation decoder across $R$ slots to detect transmitting UEs. Normalized Maximum Ratio Combining (MRC) is applied to the data at APs and the outputs are merged based on the identified hopping patterns.
  Since APs can know which UEs are active and what they transmit after $R$ slots, this scheme is suitable for delay-tolerant mMTC.
  %In Section \ref{section3_sub3}, we will introduce how to allocate pilot phase shifts, i.e., the APSP set, for each UE. These phase shifts are used to form the hopping pattern so as to improve channel estimation performance.

\subsection{APSP Set Allocation}\label{section3_sub3}
  In cell-free massive MIMO-OFDM systems, for a UE, the APs far from it cannot contribute significantly to the channel estimation performance and spatial diversity gains due to heavy path loss. Besides, the transfer of data between APs and CPU causes much fronthaul energy consumption.
  Hence, in order to improve channel estimation performance and save fronthaul energy consumption, each UE should be served by a group of APs but not all APs within the network~\cite{some_APs}.
  Hence, before allocating APSP set for each UE, AP selection is performed at CPU~\cite{b12}.
  The set of APs selected to serve the \emph{k}-th UE is denoted as $\mathcal{B}_k \subseteq \mathcal{L}$ satisfying
  $\sum_{b=0}^{\left|\mathcal{B}_k \right| - 1} \overline{\beta}_{k, b} \geq \sum_{l=0}^{L-1} \beta_{k, l} \lambda  $,
  where $\left\{ \overline{\beta}_{k, 0}, \overline{\beta}_{k, 1}, \cdots, \overline{\beta}_{k, L-1} \right\}$ includes large-scale fading coefficients between UE \emph{k} and APs in descending order, and $\lambda$ is assumed to be equal for all UEs.
  The APs belonging to $\mathcal{B}_k$ need to estimate channels and decode signals of UE $k$.
  The set of UEs served by AP \emph{l} is denoted as $\mathcal{D}_l$.

  The sporadic and independent activation of UEs and the construction of pseudo-random hopping patterns can be modeled as the process that each active UE randomly selects a pilot phase shift from its allocated set in each slot~\cite{b9}.
  We obtain the probability of having $K_a$ active UEs in \eqref{K_a}, but it is still uncertain which $K_a$ UEs are active and which phase shifts they select. To quantify these uncertainties, we employ ${{\mathcal{U}}_{{{\mathcal{K}}_{a}}}}=\left\{ \mathcal{U}_{{{\mathcal{K}}_{a}}}^{0},\mathcal{U}_{{{\mathcal{K}}_{a}}}^{1},\cdots\! ,\mathcal{U}_{{{\mathcal{K}}_{a}}}^{{{N}_{{{\mathcal{K}}_{a}}}} -1} \right\}$ to denote all possible sets of $K_a$ active UEs.
  The \emph{j}-th element in $ {{\mathcal{U}}^{i}_{{{\mathcal{K}}_{a}}}}$ is denoted as ${{\mathcal{K}}^{i,j}_{a}}$.
  In a slot, all possible choices of phase shifts for UEs in $ {{\mathcal{U}}^{i}_{{{\mathcal{K}}_{a}}}}$ are denoted as $ {{\mathcal{P}}_{i}}=\left\{ \mathcal{P}_{i}^{0},\mathcal{P}_{i}^{1}, \cdots ,\mathcal{P}_{i}^{{{N}_{\phi } -1}} \right\}$. Assuming the $p$-th choice of shifts is considered, the phase shift of UE ${{\mathcal{K}}^{i,j}_{a}}$ is denoted as $\mathcal{P}_{i,j}^{p}$.
  We assume $\nu_{k,l}$ is equal to $1$ when the \emph{l}-th AP is selected to serve the $k$-th UE and is equal to $0$ otherwise.
  Let $\boldsymbol{\Upsilon}_{k}^{\beta,0} \triangleq \begin{bmatrix}
        \nu_{k,0}\beta_{k,0}  \boldsymbol{\Upsilon}^{T}_{ k,0} &  \nu_{k,1}\beta_{ k,1}  \boldsymbol{\Upsilon}^{T}_{k,1} & \cdots & \nu_{k,L-1}\beta_{k,L-1}  \boldsymbol{\Upsilon}^{T}_{ k,L-1} \end{bmatrix}^{ T}$ and
  $\boldsymbol{\Upsilon}_{ k}^{0}  \triangleq  \begin{bmatrix}
         \nu_{k,0}\boldsymbol{\Upsilon}^{T}_{ k,0} & \nu_{k,1}\boldsymbol{\Upsilon}^{T}_{k,1} &\cdots & \nu_{k,L-1}\boldsymbol{\Upsilon}^{T}_{k,L-1} \end{bmatrix}^{T}$.
  Based on the channel estimation analysis in Section \ref{section3_sub1}, we obtain the MSE-CE of UE ${{\mathcal{K}}^{i,j}_{a}}$ averaged over its serving APs and $N_c$ subcarriers
    \begin{equation}\label{a_MSE_CE}
      \varepsilon^{0}_{ {{\mathcal{K}}^{i,j}_{a}} ,p}
      \!=\!
      {\frac{1}       {{{N}_{c}}{\left|\mathcal{B}_{ {\mathcal{K}}^{i,j}_{a}}\right|}}}
      {\sum\limits_{m=0}^{M-1}  {\sum\limits_{q=0}^{{{N}_{\mathrm{cp}}}-1}
      {\!\left\{\! {{\left[  \mathbf{\Upsilon }^{0}_{ \mathcal{K}_{a}^{i, j}} \right]}_{ m,q}}
      \!-\!\frac{{{\left[ \mathbf{\Upsilon }^{0}_{\mathcal{K}_{a}^{i,j}}\right]}_{m,q}}
      {{\left[ \mathbf{\Upsilon }_{\mathcal{K}_{a}^{i,j}}^{\beta,0} \right]}_{m,q}}}
      {\sum\limits_{j^{\prime} =0}^{K_a-1} \!\delta \!\left( \left\langle\mathcal{P}_{i,j'}^{p}\right\rangle_{\!Z}
      \!-\!\left\langle \mathcal{P}_{i,j}^{p} \right\rangle_{\!Z}\right)
      \!{{{\left[ \mathbf{\Upsilon }_{\mathcal{K}_{a}^{i,{j}^{\prime}}}^{\beta,\left\lfloor  {\mathcal{P}_{i,j'}^{p}}\!/{Z} \right\rfloor-\left\lfloor {\mathcal{P}_{i,j}^{p}}/{Z} \right\rfloor} \right]}_{m,q}}
      \!\!+\! \frac{1}{{{\rho }_{p}}Z}}}  \right\}}}}.
    \end{equation}

  We average \eqref{a_MSE_CE} over all possible sets of active UEs, all UEs in the set, and all types of phase shift selection. We obtain
    \begin{equation}\label{a_MSE_CE2}
      {{\mathbb{E}}_{\mathcal{U},{{\mathcal{K}}_{{a}}},\mathcal{P}}} (\varepsilon^{0} )
      =\sum\limits_{i=0}^{{{N}_{{{\mathcal{K}}_{a}}}-1}} {\sum\limits_{j=0}^{{{K}_{a}-1}}
      {\sum\limits_{p=0}^{{{N}_{\phi }-1}}
      {\frac{\varepsilon^{0}_{{{\mathcal{K}}^{i,j}_{a}},p}} {{{N}_{{{\mathcal{K}}_{a}}}}{{K}_{a}}{{N}_{\phi }} }}}}.
    \end{equation}
  Then we calculate the expected value of \eqref{a_MSE_CE2} accounting for the number of active UEs and obtain
    \begin{equation} \label{a_MSE_CE_0}
      \overline{\varepsilon}^{0} = \sum\limits_{K_{a}=1}^{K}{ p(K_a|K) {{\mathbb{E}}_{\mathcal{U},{{\mathcal{K}}_{{a}}},\mathcal{P}}}(\varepsilon^{0} ) } .
    \end{equation}

  The channel estimation performance suffers from pilot interference.
  In Proposition \ref{prop_2}, we show that the effect of pilot interference can be eliminated by proper APSP set allocation for UEs.
  Phase shifts in the APSP set of each UE are used to form the unique pilot phase shift hopping pattern for it.
  In correlated channels, many elements in $\mathbf{\Upsilon }_{k}^{\beta}$ are approximately $0$, which is the physical basis of the APSP set allocation scheme.
    \proposition[]\label{prop_2}
    {The minimum value of ${{\mathbb{E}}_{\mathcal{U},{{\mathcal{K}}_{{a}}},\mathcal{P}}}(\varepsilon^{0} )$ and $\overline\varepsilon ^{0}$ are given by
      \begin{equation}
        \left[{{\mathbb{E}}_{\mathcal{U},{{\mathcal{K}}_{{\!a}}},\mathcal{P}}}(\varepsilon^{0} )\right]_{\min}= {\frac{1}{K {{N}_{c}}}} \sum\limits_{k=0}^{{K-1}} {\sum\limits_{m=0}^{M-1}
        {\sum\limits_{q=0}^{{{N}_{\mathrm{cp}}}-1}
        {\frac{1}{\left|\mathcal{B}_{k}\right|}} {\left( {{\left[ \mathbf{\Upsilon }_{k}^{0}\right]}_{m,q}} - \frac{{{\left[ \mathbf{\Upsilon }_{k}^{0}\right]}_{m,q}}
        {{\left[ \mathbf{\Upsilon }_{k}^{\beta,0} \right]}_{m,q}}} {{{\left[ \mathbf{\Upsilon }_{k}^{\beta} \right]}_{m,q}} + \frac{1}{{{\rho }_{p}Z}}}  \right)}}},
      \end{equation}
      \begin{equation}
        {{{\overline{\varepsilon }}}_{\min }^{0}} = \sum\limits_{K_{a}=1}^{K}{ p(K_a|K) \left[{{\mathbb{E}}_{\mathcal{U},{{\mathcal{K}}_{{\!a}}},\mathcal{P}}}(\varepsilon^{0} )\right]_{\min} },
      \end{equation}
    and the minimum is achieved under the condition that for $\forall k,k^{\prime} \in \mathcal{K}$ and $k\neq k^{\prime}$,
      \begin{equation}\label{min_condition_0}
        \delta\left(\left\langle {{\phi }_{{{k}^{\prime}}}}\right\rangle_Z - \left\langle {{\phi }_{{{k}}}}\right\rangle_Z \right)\mathbf{\Upsilon }^{0}_{k}\odot \mathbf{\Upsilon }^{0}_{k} \odot \mathbf{\Upsilon }_{{{k}^{\prime}}}^{\beta ,\left\lfloor{{\phi }_{{{k}^{\prime}}}}/Z\right\rfloor-\left\lfloor{{\phi }_{{{k}}}}/Z\right\rfloor}=\mathbf{0}_{M\times N_{\mathrm{cp}}}.
      \end{equation}
    }
    \proof {See \ref{proof_prop_2}.}

  When APSPs are used, phase shifts are divided into $Z$ groups. There is no pilot interference for UEs using shifts in different groups.
  Channel estimation of a UE is affected by interference from other UEs in the same group exhibiting corresponding cyclic shifts in the delay domain as shown in \eqref{receive_Y2}.
  $\delta\left( \left\langle {{\phi }_{{{k}^{\prime}}}}\right\rangle_Z\! -\!\left\langle {{\phi }_{{{k}}}}\right\rangle_Z \right)\mathbf{\Upsilon }_{{{k}^{\prime}}}^{\beta ,\left\lfloor{{\phi }_{{{k}^{\prime}}}}/Z\right\rfloor-\left\lfloor{{\phi }_{k}}/Z\right\rfloor}$ is the angle-delay domain channel power spectrum interference from UE $k'$ to UE $k$ exhibiting cyclic shifts.
  By proper pilot allocation based on \eqref{min_condition_0}, there can be no overlapping between $\delta\left( \left\langle {{\phi }_{{{k}^{\prime}}}}\right\rangle_Z\! -\! \left\langle {{\phi }_{{{k}}}}\right\rangle_Z \right)\mathbf{\Upsilon }_{{{k}^{\prime}}}^{\beta ,\left\lfloor{{\phi }_{{{k}^{\prime}}}}/Z\right\rfloor-\left\lfloor{{\phi }_{k}}/Z\right\rfloor}$ and $\mathbf{\Upsilon }_{k}^{0}$, i.e., the effect of pilot interference can be mitigated and $\overline\varepsilon^{0}$ can be minimized.
  In crowded correlated cell-free massive MIMO-OFDM systems, we can allocate an APSP set for each UE to reduce ${{{\overline{\varepsilon }}}^{0}}$. This problem can be formulated as
    \begin{equation}\label{min_problem_0}
      \begin{split}
        &\min_{\mathcal{Y}_{k},k\in \mathcal{K}} \;\;\; \overline{\varepsilon }^{0}\\
        &\;\;\:{\rm{s.t.}} \;\;\;\;\; 0\leq\phi_{k,c}\leq N_cZ-1 ~~ \text{for} ~k\in \mathcal{K},~0\leq c \leq \left| \mathcal{Y} \right|-1\\
        &\:\;\;\; \;\;\;\;\;\;\;\; \phi_{k,c}\in \mathbb{Z}~~ \text{for} ~k\in \mathcal{K},~0\leq c \leq \left| \mathcal{Y} \right|-1 ,
      \end{split}
    \end{equation}
  where $\mathcal{Y}_{k}=\left\{ \phi_{k,0},\phi_{k,1},\cdots\!,\phi_{k,\left| \mathcal{Y} \right|-1 } \right\}$ is the APSP set allocated for the $k$-th UE.

  This problem is combinatorial and can be solved through an exhaustive search with large complexity. Importantly, the problem can be simplified if we aim at satisfying the condition in \eqref{min_condition_0} as much as possible. First, we define $\xi\left( \mathbf{A},\mathbf{B} \right)$ to measure the overlap degree between two non-negative matrices $\mathbf{A},\mathbf{B}$.
  When $\mathbf{A}= \mathbf{0}$ or $\mathbf{B}= \mathbf{0}$, $\xi\left( \mathbf{A},\mathbf{B} \right) \triangleq 0$.
  When $\mathbf{A},\mathbf{B}\neq \mathbf{0}$, we have
  \begin{equation}\label{overlap}
    \xi\left( \mathbf{A},\mathbf{B} \right)\triangleq \frac{\left| \sum\nolimits_{i,j}{{{\left[ \mathbf{A}\odot \mathbf{B} \right]}_{i,j}}} \right|}{\sqrt{\sum\nolimits_{i,j}{\left[ \mathbf{A} \right]_{i,j}^{2}}}\cdot \sqrt{\sum\nolimits_{i,j}{\left[ \mathbf{B} \right]_{i,j}^{2}}}}\in \left[ 0,1 \right].
  \end{equation}

  In Algorithm \ref{alg:1}, k-means clustering method is utilized to partition $K$ UEs into $J$ clusters. Next, APSP set is allocated for each UE to make the effect of pilot interference from other UEs in the same cluster as small as possible.
  Algorithm \ref{alg:1} should be performed before pilot transmission, i.e., without the knowledge of UEs' activity.
  Hence, it is computed at CPU assuming all UEs are active and their statistical CSI is known. Hence, this scheme is insensitive to the active pattern of UEs and does not need to be recomputed frequently.

  \begin{algorithm}
    \footnotesize
    \caption{APSP Set Allocation Algorithm}
    \label{alg:1}
    \begin{algorithmic}[1]

    \REQUIRE
    The angle-delay domain channel power spectrum $\{\mathbf{\Upsilon}_k:k\in\mathcal{K}\}$;
    the threshold $\lambda$ and $\gamma$;
    the phase shift set $\mathbf{\Psi}$;
    the large-scale fading $\boldsymbol{\beta}_k \triangleq  \left[\beta_{k,0}\!\;\;\beta_{k,1}\cdots \beta_{k,L-1} \right]$ for $\forall k \in \mathcal{K}$.\\
    \ENSURE
    The APSP set allocated to each UE $\left\{ \mathcal{Y}_k, k\in \mathcal{K}\right\}$.\\

    % k-means 分组
    \STATE {Centroids are randomly chosen as $\tilde{\boldsymbol{\beta}}_{j}\in \mathbb{C}^{L\times 1}$ for $j=0,1,\cdots,J-1$}
    \STATE {For $\forall k \in \mathcal{K}$, UE $k$ is assigned to the cluster with $\max\limits_{j=0,1,\cdots,J-1}\xi\left(\boldsymbol{\beta}_k, \tilde{\boldsymbol{\beta}}_{j}\right)$}
    \STATE {Centroids are updated by averaging over UEs belonging to respective clusters and UEs are reassigned until the assignments no longer change. Denote $\mathcal{C}_j$ as the set of UEs belonging to the $j$-th cluster}

    % APSP set 分配
    \FOR {$j=0,1,\cdots,J-1$}
    \STATE {$\left|\mathcal{Y} \right|$ shifts are randomly chosen to form the set $\mathcal{Y}_{\mathcal{C}_j(0)}$ allocated for UE $\mathcal{C}_j(0)$.
    Initialize the allocated UE set $\mathcal{K}^{\rm{al}}_{j}=\{\mathcal{C}_j(0)\}$ and the unallocated UE set $\mathcal{K}^{\rm{un}}_{j}=\mathcal{C}_j\backslash\{\mathcal{C}_j(0)\}$. }

    \FOR {$k\in \mathcal{K}^{\mathrm{un}}_{j}$}
    \STATE {Search for $\left|\mathcal{Y}\right|$ pilot phase shifts $\phi\in\mathbf{\Psi}$ to form the APSP set $\mathcal{Y}_k$ allocated for the $k$-th UE, which should  satisfy
    $ \sum\limits_{k^{\prime}\in\mathcal{K}^{\mathrm{al}}_{j}}  {\frac{1}{\left|\mathcal{K}^{\mathrm{al}}_{j} \right|}} \mathop {\max}\limits_{\phi_{k^{\prime}}\in \mathcal{Y}_{k^{\prime}}} \left\{\delta \left( \left\langle {{\phi }_{{{k}^{\prime}}}}\right\rangle_{Z} -\left\langle {{\phi }}\right\rangle_{ Z} \right) \xi \left( \mathbf{\Upsilon}_k^0 \odot \mathbf{\Upsilon}_k^0, \mathbf{\Upsilon}_{k^{\prime}}^{\beta,\left\lfloor{{\phi }_{{{k}^{\prime}}}}/Z\right\rfloor-\left\lfloor{{\phi }}/Z\right\rfloor} \right) \right\}\leq \gamma$}

    \STATE {If $\left|\mathcal{Y}\right|$ pilot phase shifts are not found in step 7,
    then search for $\left|\mathcal{Y}\right|$ shifts from $\mathbf{\Psi}$ corresponding to the $\left|\mathcal{Y}\right|$ smallest
    $ \sum\limits_{k^{\prime}\in\mathcal{K}^{\mathrm{al}}_{j}}  \mathop {\max}\limits_{\phi_{k^{\prime}}\in \mathcal{Y}_{k^{\prime}}} \left\{\delta\left( \left\langle {{\phi }_{{{k}^{\prime}}}}\right\rangle_{Z} - \left\langle {{\phi }}\right\rangle_{ Z} \right) \xi\left(  \mathbf{\Upsilon}_k^0 \odot  \mathbf{\Upsilon}_k^0, \mathbf{\Upsilon}_{k^{\prime}}^{\beta,\left\lfloor{{\phi }_{{{k}^{\prime}}}}/Z\right\rfloor-\left\lfloor{{\phi }}/Z\right\rfloor} \right) \!\right\}$}

    \STATE {
    Update $\mathcal{K}^{\rm{un}}_{j}:=\mathcal{K}^{\rm{un}}_{j}\backslash\{k\},
    \mathcal{K}^{\rm{al}}_{j}:=\mathcal{K}^{\rm{al}}_{j}\cup\{k\}$}
    \ENDFOR
    \ENDFOR
    \end{algorithmic}
    \end{algorithm}

  We evaluate the complexity of the proposed scheme as follows.
  The complexity of k-means clustering is $\mathcal{O}\left( eKJL\right)$, where $e$ is the number of iterations needed until convergence and is often small~\cite{b25}.
  $\gamma$ is the overlap degree threshold of two matrices, which can balance the complexity and performance of the APSP set allocation scheme.
  To search for $\left|\mathcal{Y} \right|$ phase shifts for each UE, no more than $N_c\left|\mathcal{Y}\right| \frac{K'(K'-1)}{2}$ calculations of \eqref{overlap} are needed assuming the number of UEs in a cluster is approximately $K' = \lfloor \frac{K}{J}\rfloor$, since there is no need to calculate \eqref{overlap} when $\delta\left( \left\langle {{\phi }_{{{k}^{\prime}}}}\right\rangle_{Z} -\left\langle {{\phi }}\right\rangle_{Z} \right)=0$.
  For each calculation, considering $N_{\mathrm{cp}}$ is relatively small, the maximal scalar multiplication number for calculating $\mathbf{\Upsilon}_k^0 \odot \mathbf{\Upsilon}_k^0$ is $\mathcal{O}\left( M \right)$, the complexity for obtaining $\mathbf{\Upsilon}_{k^{\prime}}^{\beta,\left\lfloor\phi_{k^{\prime}}/Z\right\rfloor-\left\lfloor\phi/Z\right\rfloor}$ based on \eqref{p_shift} is neglected because it only needs cyclic column shift and truncation,
  and the scalar multiplication number for calculating $\xi\left(\mathbf{\Upsilon}_k^0 \odot \mathbf{\Upsilon}_k^0, \mathbf{\Upsilon}_{k^{\prime}}^{\beta,\left\lfloor\phi_{k^{\prime}}/Z\right\rfloor-\left\lfloor\phi/Z\right\rfloor}\right)$ is $\mathcal{O}\left( M \right)$ when $\mathbf{\Upsilon}_k^0 \odot \mathbf{\Upsilon}_k^0$ and $\mathbf{\Upsilon}_{k^{\prime}}^{\beta,\left\lfloor\phi_{k^{\prime}}/Z\right\rfloor-\left\lfloor\phi/Z\right\rfloor}$ are figured out.
  Hence, the computational complexity of allocating APSP sets for all UEs is no more than $\mathcal{O}\left(J M^{2} K'^{2}N_c \right)$.
  When $\lambda$ is small enough, it can be reduced to $\mathcal{O}\left( JN^{2} K'^{2}N_c \right)$.
  When $\gamma$ is large enough, it can be further reduced to $\mathcal{O}\left( JN^{2} K'^{2} \right)$.
  Compared with the scheme allocating APSP sets for all UEs directly, the complexity of the proposed scheme is greatly reduced since UEs are partitioned via k-means method with liner complexity, which makes the proposed scheme can be utilized in crowded scenarios.

\section{Spectral Efficiency and Power Control} \label{section4}
  In this section, we analyse a lower bound of uplink SE in cell-free massive MIMO-OFDM systems. Then we propose a power control scheme to maximize the minimum SE lower bound among active UEs.
\subsection{A Lower Bound of SE}\label{section4_sub1}
  In the uplink data transmission phase, $K_a$ active UEs simultaneously transmit data to APs. The data of UE \emph{k} over the \emph{s}-th subcarrier in an OFDM symbol is denoted by $x_{k,s}\sim \mathcal{CN}(0,1)$, which is weighted by a power control coefficient $\sqrt{\eta_{k}}$ satisfying $0 \leq \eta_{k} \leq 1$.
  The received signal at AP $\emph{l}$ over the \emph{s}-th subcarrier is given by
    \begin{equation}
      \mathbf{y}_{u,l,s}=\sqrt{\rho_{u}} \sum_{k^{\prime} \in \mathcal{K}_{a}} \sqrt{\eta_{k' }} \mathbf{g}_{k',l,s}^{\beta} {x}_{k',s}+\mathbf{w}_{l,s} \in \mathbb{C}^{N \times 1},
    \end{equation}
  where $\rho_{u}$ is the normalized uplink SNR and $\mathbf{w}_{l,s}$ is the AWGN vector at the \emph{l}-th AP over the \emph{s}-th subcarrier including i.i.d. $\mathcal{CN} \;\!(0,1)$ elements.
  The \emph{l}-th AP needs to detect the symbols transmitted from active UEs belonging to the set $\mathcal{D}_l$. The normalized MRC is utilized, i.e., $\mathbf{c}_{k, l,s}^{\beta}=\hat{\mathbf{g}}_{k, l,s}^{\beta}/{ \left\|\hat{\mathbf{g}}_{k, l,s}^{\beta}\right\|} ^{2}$, where $\hat{\mathbf{g}}_{k,l,s}^{\beta}$ is the estimated channel response vector between the \emph{k}-th UE and the \emph{l}-th AP over the \emph{s}-th subcarrier.
  The symbol of UE $k$ over the \emph{s}-th subcarrier detected by the \emph{l}-th AP is given by
    \begin{align}
      {{r}_{k,l,s}}&=\left(\mathbf{c}_{k,l,s}^{\beta }\right)^{H} {{\mathbf{y}}_{u,l,s}} \notag \\
      &=\sqrt{{{\rho }_{u}}}\sqrt{{{\eta }_{k}}} \left(\mathbf{c}_{k,l,s}^{\beta }\right)^{\!H} {\mathbf{g}}_{k,l,s}^{\beta }{{x}_{k,s}}
      +\left(\mathbf{c}_{k,l,s}^{\beta }\right)^{\!H} \left( \sqrt{{{\rho }_{u}}}\!\sum\limits_{{k}'\in \mathcal{K}_{a} \backslash \left\{ k \right\}}\! {\sqrt{{{\eta }_{{{k}'}}}}\mathbf{g}_{{k}',l,s}^{\beta }{{x}_{{k}',s}}} \right)
      +\left(\mathbf{c}_{k,l,s}^{\beta }\right)^{\!H}  {{\mathbf{w}}_{l,s}}.
    \end{align}

  Next, each AP sends the detected symbols to the CPU via a fronthaul network. For UE \emph{k}, the CPU only sees the detected symbols from APs serving it, which can be represented as
    \begin{equation}\label{r_ks}
      {{r}_{k,s}}=\sum\limits_{l=0}^{L-1}{\nu_{k,l}{{r}_{k,l,s}}}
      = \sqrt{{{\rho }_{u}}}\sum\limits_{{k}'\in \mathcal{K}_{a} } {\sqrt{{{\eta }_{{{k}'}}}}\left( \mathbf{c}_{k,s}^{\beta,0 }\right)^{ H} \mathbf{g}_{{k}',s}^{\beta }{{x}_{{k}',s}}}
      +\left(\mathbf{c}_{k,s}^{\beta,0 }\right)^{ H} {{\mathbf{w}}_{s}},
    \end{equation}
  where $\mathbf{c}_{k,s}^{\beta,0}  \triangleq  \operatorname{vec} \left\{
         \nu_{k,0}\mathbf{c}_{k,0,s}^{\beta}, \nu_{k,1}\mathbf{c}_{k,1,s}^{\beta}, \cdots  , \nu_{k,L-1}\mathbf{c}_{k,L-1,s}^{\beta} \right\}$, and $\mathbf{w}_{s}  \triangleq   \operatorname{vec} \left\{ \mathbf{w}_{0,s}, \mathbf{w}_{1,s}, \cdots  , \mathbf{w}_{L-1,s} \right\}$.

  Define $\varpi\triangleq\frac{{{N}_{c}}}{{{N}_{c}}+{{N}_{\mathrm{cp}}}}\frac{Z_a-Z}{Z_a}$.
  Based on the use-and-then-forget method~\cite{b27}, a lower bound of SE of UE \emph{k} over the \emph{s}-th subcarrier is given in \eqref{SE_lb} with $\text{SINR}_{k,s}^{\mathrm{lb}}$ given in \eqref{SINR_lb} (see \ref{proof_prop_3} for derivations).
    \begin{equation}\label{SE_lb}
      \mathrm{SE}_{k, s}^{\mathrm{lb}}\left( \left\{ {{\eta }_{k}} \right\} \right)=\varpi
      \log _{2}
      \left\{1+\text{SINR}_{k,s}^{\mathrm{lb}}\left( \left\{ {{\eta }_{k}} \right\} \right)\right\},
    \end{equation}
    \begin{equation}\label{SINR_lb}
      \text{SINR}_{k,s}^{\mathrm{lb}} \left( \left\{ {{\eta }_{k}} \right\} \right)=
      \frac{\eta_{k}\left|\mathbb{E}\left\{\left(\mathbf{c}_{k,s}^{\beta,0}\right)^{H} \mathbf{g}_{k, s}^{\beta}\right\}\right|^{2}}
      {\sum\limits_{k'\in \mathcal{K}_{a}} \eta_{k'}\mathbb{E} \left\{\left|\left(\mathbf{c}_{k,s}^{\beta,0}\right)^{ H}  \mathbf{g}_{k' , s}^{\beta}\right|^{2}\right\}
      -\eta_{k}\left|\mathbb{E} \left\{\left(\mathbf{c}_{k,s}^{\beta,0}\right)^{ H}  \mathbf{g}_{k, s}^{\beta}\right\}\right|^{2}
      +\frac{1}{\rho_{u}} \mathbb{E} \left\{\left\| \mathbf{c}_{k,s}^{\beta,0}\right\|^{2}\right\} }.
    \end{equation}
  In correlated channels with less hardening, the lower bound is tighter if the normalized MRC is used, compared with the lower bound in~\cite{b4,b10} using MRC as $\mathbf{c}_{k, l,s}^{\beta}=\hat{\mathbf{g}}_{k, l,s}^{\beta}$. The reason and comparison are shown in~\cite{b26}.
  In this work, since the correlated channels in cell-free massive MIMO systems are considered, we utilize the tighter normalized MRC.
  Considering the closed-form expression of SE lower bound is tricky to be derived for normalized MRC in correlated channels, we adopt Monte Carlo simulation to obtain each expectation over the random channel realizations in~\eqref{SINR_lb}.
  In Section~\ref{section4_sub2}, we optimize the power coefficients based on~\eqref{SE_lb}.
  The reason why we adopt a lower bound of SE to optimize power coefficients is that the expression of SE is not analytically tractable, though it does not rely on channel hardening and can reflect the real SE even in correlated channels if the transmitted signals are $\mathcal{CN} \;\!(0,1)$ RVs~\cite{b26}.
  For simulation comparison in Section~\ref{section5}, the expression of SE is also given here~\cite{b26}
  \begin{equation}\label{SE}
      {{\text{SE}}_{k,s}}\!\left( \left\{ {{\eta }_{k}} \right\} \right)
      \!=\!\varpi
      \mathbb{E} \left\{ \!{{\log }_{2}}\left\{
      1 +
      \frac{{{\rho }_{u}}{{\eta }_{k}}{{\left| \left(\mathbf{c}_{k,s}^{\beta,0}\right)^{H} \hat{\mathbf{g}}_{k,s}^{\beta } \right|}^{2}}}
      { \left(\mathbf{c}_{k,s}^{\beta,0} \right)^{ \!H}
      \!\left(
      \!{{\rho }_{u}} \sum\limits_{{{k}'\in \mathcal{K}_{a}  \backslash \left\{ k \right\}} }\!{{ {\eta }_{{{k}'}}}{{ \hat{\mathbf{g}}_{{k}',s}^{\beta } \left(\hat{\mathbf{g}}_{{k}' ,s}^{\beta }\right)^{ \!H}}}}
      \!\!+\!{{\rho }_{u}} \sum\limits_{{k}'\in {{\mathcal{K}}_{a}}}{ \!{{\eta }_{{k}'}} {{\mathbf{R}}_{\mathbf{\tilde{g}}_{_{{k}' ,s}}^{\beta}}}}
      \!\!+\! \mathbf{I}_{M}  \!\right) \mathbf{c}_{k,s}^{\beta,0}} \right\}\right\},
    \end{equation}
  where the expectation is with respect to channel estimates, and ${{\mathbf{R}}_{\mathbf{\tilde{g}}_{_{{k}' ,s}}^{\beta}}}$ is represented as
    \begin{equation}\label{R_error}
      {{\mathbf{R}}_{\mathbf{\tilde{g}}_{_{{k}',s}}^{\beta}}} =  \frac{1}{{{N}_{c}}}{{\mathbf{V}}_{ M\times M}}
       \operatorname{diag} \left\{ \begin{matrix}
      \operatorname{sum} \left\{  {{\left[ {{{\mathbf{\Xi}}}_{k'}^{\beta}} \right]}_{0,:}}  \right\} & \operatorname{sum}\left\{ {{\left[ {{{\mathbf{\Xi}}}_{k'}^{\beta}} \right]}_{1,:}} \right\}& \cdots   &
      \operatorname{sum} \left\{  {{\left[ {{{\mathbf{\Xi}}}_{k'}^{\beta}}  \right]}_{M-1,:}} \right\}
      \end{matrix}  \right\} {{\mathbf{V}}_{M\times M}^{H}}.
    \end{equation}

\subsection{Power Control}\label{section4_sub2}
  A wireless system should provide good service to all UEs. The minimum data rate among UEs is an important indicator of system performance. In this subsection, we aim to maximize the minimum uplink SE lower bound among active UEs using max-min power control, and hence, improve the real minimum SE among active UEs in correlated cell-free massive MIMO-OFDM systems.
  The problem of maximizing the minimum uplink SE lower bound among active UEs can be formulated as
    \begin{equation}\label{max-min_problem_1}
      \begin{split}
        &\max_{\left\{\eta_{k} \right\}}\; \min_{k\in \mathcal{K}_{a}}\; \text{SE}_{k,s}^{\mathrm{lb}} \left( \left\{ {{\eta }_{k}} \right\} \right)\\
        &\;\;\!{\rm{s.t.}} \;\;\;\; 0\leq\eta_{k}\leq 1 , ~{k\in \mathcal{K}_{a}}.
      \end{split}
    \end{equation}

  Problem \eqref{max-min_problem_1} can be equivalently reformulated as
    \begin{equation}\label{max-min_problem_2}
      \begin{split}
        &\max_{\left\{\eta_{k} \right\},t}\;\; t \\
        &\;\;\!{\rm{s.t.}}\;\;\; t \leq \text{SINR}_{k,s}^{\mathrm{lb}}\left( \left\{ {{\eta }_{k}} \right\} \right) , ~{k\in \mathcal{K}_{a}}\\
        &\;\;\;\;\;\;\;\;\; 0\leq\eta_{k}\leq 1 , ~{k\in \mathcal{K}_{a}}.
      \end{split}
    \end{equation}

  Problem \eqref{max-min_problem_2} is quasilinear because inequalities are linear for a given \emph{t}. It can be solved using bisection as shown in~\cite{b4,b11}, where initial values of $t_{\min}$ and $t_{\max}$ are chosen and the range is bisected until it can be accepted.
  However, it is not easy to choose proper initial value $t_{\max}$ in practice. This is because a larger value can increase the iteration times and a smaller value cannot satisfy the requirement.
  Hence, we utilize Dinkelbach's Algorithm to efficiently solve \eqref{max-min_problem_2} as shown in Algorithm \ref{alg:2}, where only the initial value of $t_{\min}$ is needed. We can obtain the globally optimal solution of this problem by iteratively solving linear programs in \eqref{max-min_problem_3} and updating $t_{\min}$.
    \begin{equation}\label{max-min_problem_3}
      \begin{split}
        &\max_{\left\{\eta_{k} \right\}}\; w \\
        &\;\;\!{\rm{s.t.}}\; \;\; {\eta_{k}
        \!\left( 1\!+\!t_{\min}\right)\!
        \left|\mathbb{E}\!\left\{\!\left(\!\mathbf{c}_{k,s}^{\beta,0}\!\right)^{\!\!H} \!\!\mathbf{g}_{k, s}^{\beta}\!\right\}\right|^{2} }\!\!-t_{\min} \!\left(
        {\sum\limits_{k'\in \mathcal{K}_{a} }\! \!\eta_{k'}\mathbb{E} \left\{\!\left|\left(\!\mathbf{c}_{k,s}^{\beta,0}\!\right)^{\!\!H}\!\! \! \mathbf{g}_{k'\!, s}^{\beta}\right|^{2}\!\right\}
        \!+\frac{1}{\rho_{\!u}} \mathbb{E}\!\left\{\left\| \mathbf{c}_{k,s}^{\beta,0}\right\|^{2}\!\right\} } \!\!\right)\!\!  \geq\!   w, ~{k\!\in \mathcal{K}_{a}} \\
        &\;\;\;\;\;\;\;\;\; 0\leq\eta_{k}\leq 1 , ~{k\in \mathcal{K}_{a}}.
      \end{split}
    \end{equation}
    \begin{algorithm}
    \footnotesize
    \caption{Dinkelbach's Algorithm for Solving \eqref{max-min_problem_1}}
    \label{alg:2}
    \begin{algorithmic}[1]
    \STATE {Initialize $t_{\min}=0$ satisfying the constraint in \eqref{max-min_problem_2}, and choose a tolerance $\epsilon$.}
    \STATE {
    Solve the linear problem in \eqref{max-min_problem_3} and obtain the optimal solution $\left\{\eta_k^{\ast} \right\}$ and the optimal value $w^{\ast}$, where the expectations over the random channel realizations can be computed separately by means of Monte Carlo simulation. }
    \STATE {Update $t_{\min} := \min\limits_{k\in \mathcal{K}_{a}}\left\{ \text{SINR}_{k,s}^{\mathrm{lb}}\left( \left\{ {{\eta }_{k}^{\ast}} \right\}\right)\right\}$.}
    \STATE {Stop if $w^{\ast}\leq \epsilon$. Otherwise, go to Step 2.}
    \end{algorithmic}
    \end{algorithm}

    The Dinkelbach's Algorithm always converges superlinearly and often (locally) quadratically~\cite{b28}.
    The number of iterations is assumed to be $T_D$.
    For each iteration, when the interior-point method is adopted, the linear problem in step 2 can be solved with polynomial complexity $\mathcal{O}\left( K_a^{3.5} L_{b}^2\right)$~\cite{b29}.
    In this case, the number of bits in the input satisfies $ L_{b} =\sum\limits_{k\in \mathcal{K}_{a} } {\left\lceil 1 + \log_2\left(1 + \left( 1+t_{\min}\right)       \left|\mathbb{E}\left\{\left(\mathbf{c}_{k,s}^{\beta,0}\right)^{H}\ \mathbf{g}_{k, s}^{\beta}\right\}\right|^{2} \right) \right\rceil}
    + \sum\limits_{k,k'\in \mathcal{K}_{a} } {\left\lceil 1 +\log_2\left(1+ t_{\min} \mathbb{E} \left\{\left| \left(\mathbf{c}_{k,s}^{\beta,0} \right)^{ H}  \mathbf{g}_{k' , s}^{\beta}\right|^{2}\right\}   \right) \right\rceil}
    + \sum\limits_{k \in \mathcal{K}_{a}} {\left\lceil 1+\log_2\left(1+\frac{t_{\min}}{\rho_{ u}} \mathbb{E} \left\{ \left\| \mathbf{c}_{k,s}^{\beta,0}\right\|^{2} \right\} \right) \right\rceil}
    +6K_a  $.
    Step 3 is performed with complexity $\mathcal{O}\left( K_a^2\right)$.
    Hence, the complexity of Algorithm \ref{alg:2} is $\mathcal{O}\left( K_a^{3.5} L_{b}^2 T_D\right)$, i.e., a polynomial of the number of active UEs, which is usually far less than that of UEs in crowded scenarios.
    Algorithm \ref{alg:2} should be recomputed at CPU if  statistical CSI or activity pattern is changed.

  In practice, it should be performed after UE detection in two ways depending on specific demands.
  If data needs to be decoded in a short time, active UEs should transmit pilots and data with full power in the first uplink frame shown in Fig.~\ref{fig:2}. Once APs receive these pilots, active UEs can be detected based on hopping patterns and power coefficients can be optimized and sent to UEs during downlink transmission. Next, pilots can be  transmitted with full power and data can be transmitted with power control.
  If the minimum SE among active UEs needs to be improved, the first $ZR$ OFDM symbols can be used to transmit pilots for UE detection. Then power coefficients is optimized at CPU and  transmitted to UEs. Finally, pilots are transmitted with full power and data is transmitted with power control. These pilots are for UE detection and channel estimation.

\section{Numerical Results and Discussions} \label{section5}
\subsection{Large-Scale Fading Model and System Parameters}\label{section5_sub1}
  We assume that APs and UEs are independently and uniformly distributed within a square of size $1 \times 1 ~\mathrm{km}^{2}$.
  Each AP is equipped with two 100-antenna ULAs. The large-scale fading coefficient $\beta_{k,l}$ between the \emph{k}-th UE and the \emph{l}-th AP is modeled as $10\log_{10}\left(\beta_{k,l}\right) = \mathrm{PL}_{k,l} \mathrm{SH}_{k,l}$,
  where the shadow fading $\mathrm{SH}_{k,l}\sim  \mathcal{N} \left(0,\sigma_{\mathrm{sh}}^{2} \right)$ with $\sigma_{\mathrm{sh}} = 8 ~\mathrm{dB}$. Denote by $d_{k,l}$ the distance between UE \emph{k} and AP \emph{l}. If $d_{k,l}\leq d_1$, there is no shadowing.
  We model the three-slope path loss $\mathrm{PL}_{k,l}$ as~\cite{large_scale_fading}
    \begin{equation}
      \mathrm{PL}_{k, l}=\left\{\begin{array}{c}
      {-\chi-35 \log _{10}\left(d_{k, l}{\!~/\!\!~1\!~\rm m}\right), \text { if } d_{k, l}>d_{1}}
      \\ {-\chi-15 \log _{10}\left(d_{1}{\!~/\!\!~1\!~\rm m}\right)-20 \log _{10}\left(d_{k, l}{\!~/\!\!~1\!~\rm m}\right),} {\text { if } d_{0}<d_{k, l} \leq d_{1}}
      \\ {-\chi-15 \log _{10}\left(d_{1}{\!~/\!\!~1\!~\rm m}\right)-20 \log _{10}\left(d_{0}{\!~/\!\!~1\!~\rm m}\right), \text { if } d_{k, l} \leq d_{0}}\end{array}\right. ,
    \end{equation}
  where $\chi \triangleq 46.3+33.9 \log _{10}(f_c{\!~/\!~1\!~\rm {MHz}})-13.82 \log _{10}\left(h_{\mathrm{AP}}{\!~/\!~1\!~\rm m}\right) -\left(1.1 \log _{10}(f_c{\!~/\!~1\!~\rm {MHz}})-0.7\right) h_{\mathrm{u}}+\left(1.56 \log _{10}(f_c{{\!~/\!~1\!~\rm {MHz}}})-0.8\right)$,
  $f_c$ is the carrier frequency (in MHz), and $h_{\mathrm{AP}}$ and $h_{\mathrm{u}}$ are the antenna height (in m) of AP and UE, respectively. The values of major parameters are given in Table \ref{Tab:1}.

  The noise power is $\sigma_{w}^{2}\!=\!B \times k_{\mathrm{B}} \times T_{0} \times NF~\!(\mathrm{W})$, where $T_{0} = 290 ~\mathrm{(Kelvin)}$ is the noise temperature, $k_{\mathrm{B}}\!=\!1.381\times 10^{-23}~\mathrm{(Joule ~per ~Kelvin)}$ is the Boltzmann constant, and $NF=9~\mathrm{dB}$ is the noise figure~\cite{b11}. The corresponding normalized transmitting SNRs in the training phase and data transmission phase are assumed to be equal, which is computed by dividing $P_x$ by $\sigma_{w}^{2}$.

  We consider channels with 30 taps in the delay domain, unless stated otherwise.
  The channel power from a UE to an AP is normalized as $\sum_{i, j}\left[\mathbf{\Upsilon}_{k,l}\right]_{i, j}\!=\!N N_{\mathrm{cp}}$.
  \begin{table}[!t]
  \footnotesize
    \caption{System Parameters for the Simulation} \label{Tab:1}
    \renewcommand{\arraystretch}{1.2}
    \centering
    \begin{tabular}{|c|c|}
      \hline
      Parameter & Value\\ \hline
      Bandwidth B & 20 MHz\\  \hline
      Carrier frequency $f_c$ & 2 GHz\\ \hline
      Sampling duration $T_s$& 48.8 ns\\ \hline
      Subcarrier number $N_c$, Guard interval $N_{\mathrm{cp}}$ & 1024, 144\\  \hline
      the number of UEs $K$, the number of clusters $J$ & $200,2$\\  \hline
      the number of OFDM symbols in a slot $Z_{a}$ & $7$\\  \hline
      $h_{\mathrm{AP}}$, $h_{\mathrm{u}}$, $d_1$, $d_0$ & 15, 1.65, 50, 10 m\\  \hline
      Maximum transmitting power $P_x$& $0.5 $ W\\  \hline
      Pilot phase shift number per UE $\left|\mathcal{Y} \right|$ & 4\\   \hline
      $\gamma$ & $10^{-8} (Z=1), 10^{-13}(Z=2)$ \\  \hline
      $\iota_k$ & 0.4\\   \hline
      $P_{\mathrm{ip,} k}$, $P_{\mathrm{ip,} l}$, $P_{0, l}$& 0.1, 0.1, 0.825 W\\   \hline
      $P_{\mathrm{bt,} l}$ & 0.25 W/(Gbits/s)\\
      \hline
    \end{tabular}
    \end{table}

\subsection{Results and Discussions}\label{section5_sub2}
\subsubsection{The MSE-CE Performance with Different Numbers of Active UEs and Pilot Symbols}
\begin{figure*}[!t]
\centering
\subfloat[]{\includegraphics[width=0.49\linewidth]{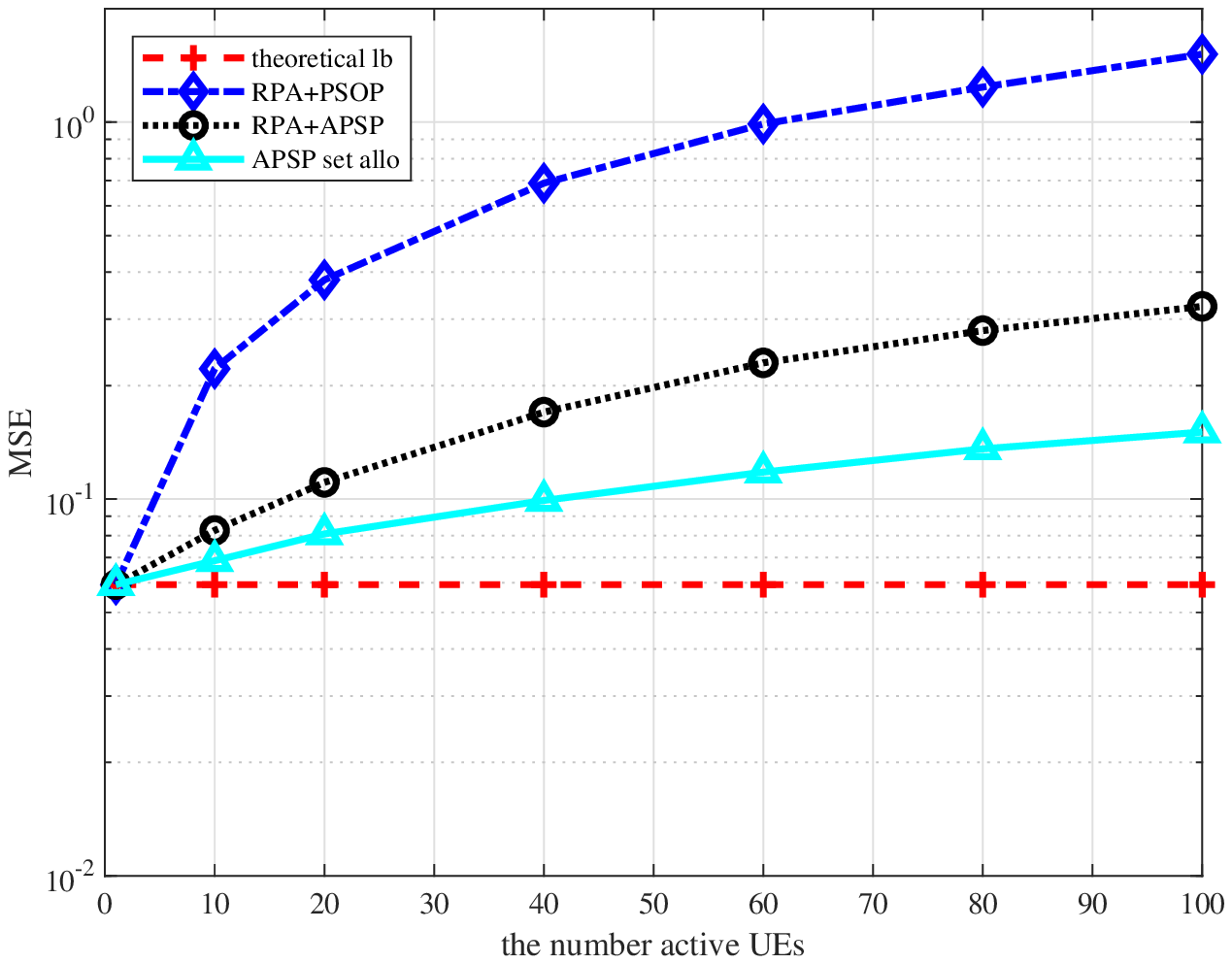}\label{fig:3a}}
\hfil
\subfloat[]{\includegraphics[width=0.49\linewidth]{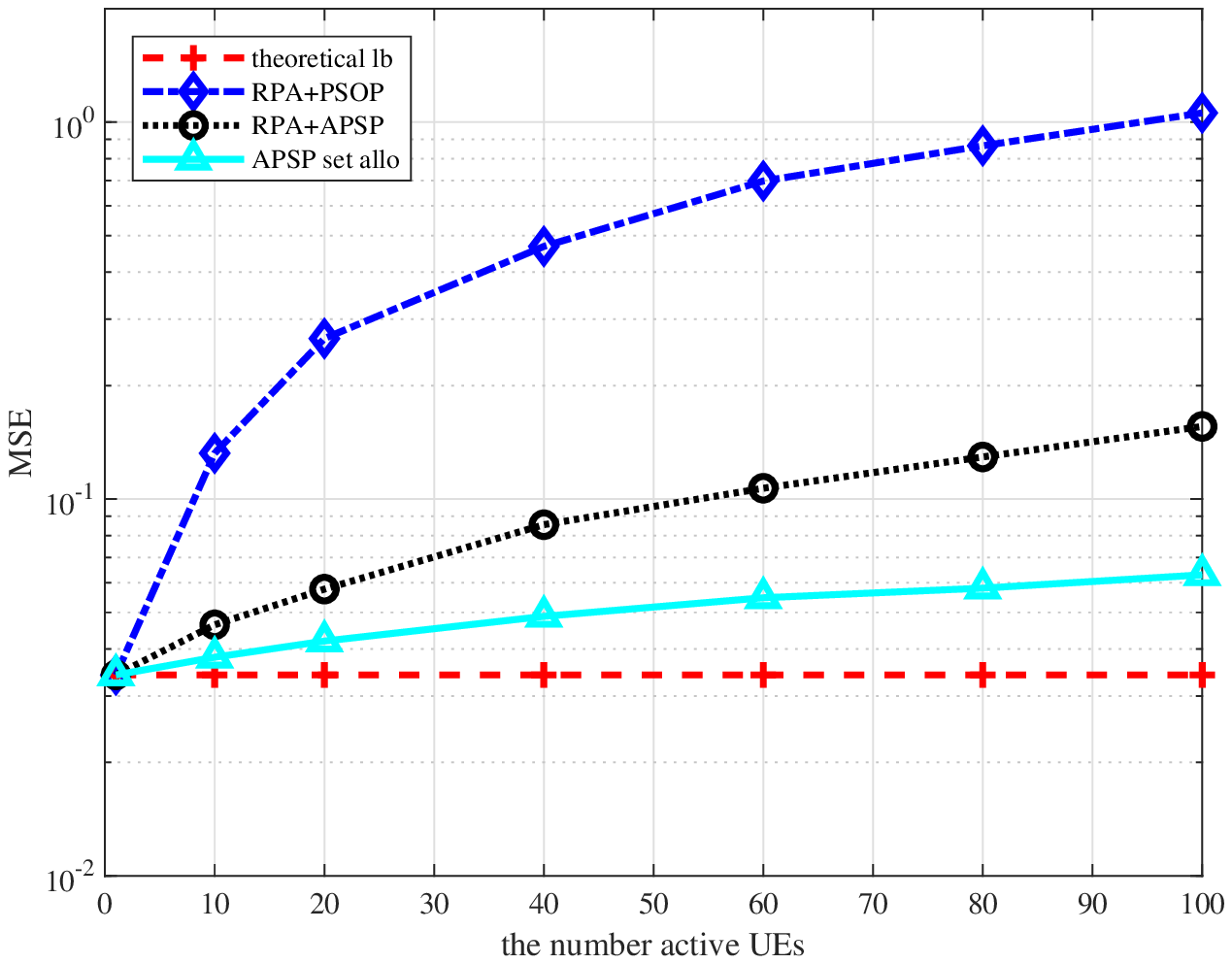}\label{fig:3b}}
\caption{Comparison of MSE-CE of pilot assignment schemes and theoretical lower bound with AP selection versus different numbers of active UEs assuming $\zeta\! =\! 0.2~\mu \mathrm{s}$, $\varsigma \!= \!2^\circ$, $L\!=\!10$, and $\lambda\!=\!0.7$. (a) Results are shown with $Z\!=\!1$. (b) Results are shown with $Z\!=\!2$.}
\label{fig:3}
\end{figure*}

  In Figure~\ref{fig:3}, we compare the MSE-CE performance of pilot assignment schemes and theoretical lower bound versus different numbers of active UEs.
  In the legend, we rewrite ``lower bound'' and ``allocation'' to ``lb'' and ``allo'' for short, respectively.
  Figure~\ref{fig:3a} and Figure~\ref{fig:3b} show the performance with $Z=1$ and $Z=2$ OFDM symbols used for pilot transmission in a slot, respectively.
  It is shown that the performance of each scheme is improved when more OFDM symbols are used for pilot transmission.
  For the PSOP-based RPA scheme, UEs are randomly assigned pilots from phase shift orthogonal pilot set. The number of PSOPs is $\lfloor ZN_c/N_{cp}\rfloor$.
  It can cause serious pilot interference, since the number of PSOPs is far less than that of UEs.
  For the APSP-based RPA scheme, UEs are randomly assigned phase shifts from $\mathbf{\Psi}$. It can utilize channel sparsity and outperform the PSOP-based scheme.
  Although the proposed APSP set allocation scheme is suboptimal in general compared with exhaustive search, substantial performance gains in terms of MSE-CE can still be achieved for different numbers of active UEs and pilot symbols. This is because it can reduce the overlap of angle-delay domain channel power distributions between the desired UE and interference with corresponding cyclic shifts in the delay domain.
  As observed from the trend of each curve, the performance gain of the proposed scheme can still be obvious when the number of active UEs is more than 100.
  Moreover, the theoretical lower bound is plotted, where each UE is assumed to be assigned a dedicated orthogonal pilot.
  As the number of active UEs increases, the MSE-CE of each pilot assignment scheme and the gap between it and corresponding lower bound increase.

\subsubsection{The MSE-CE Performance with Different Channel Sparsity}
\begin{figure*}[!t]
\centering
\subfloat[]{\includegraphics[width=0.49\linewidth]{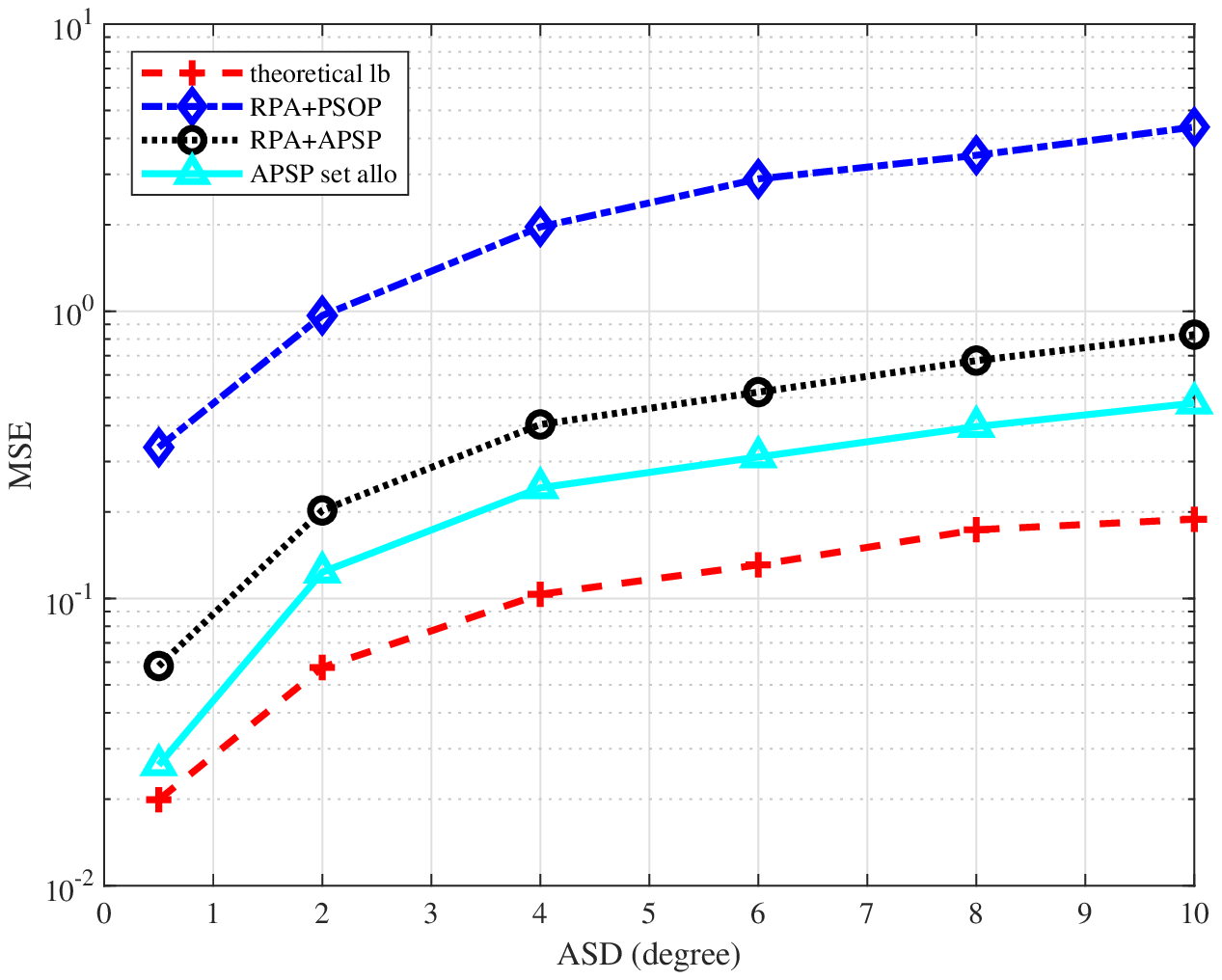}\label{fig:4a}}
\hfil
\subfloat[]{\includegraphics[width=0.49\linewidth]{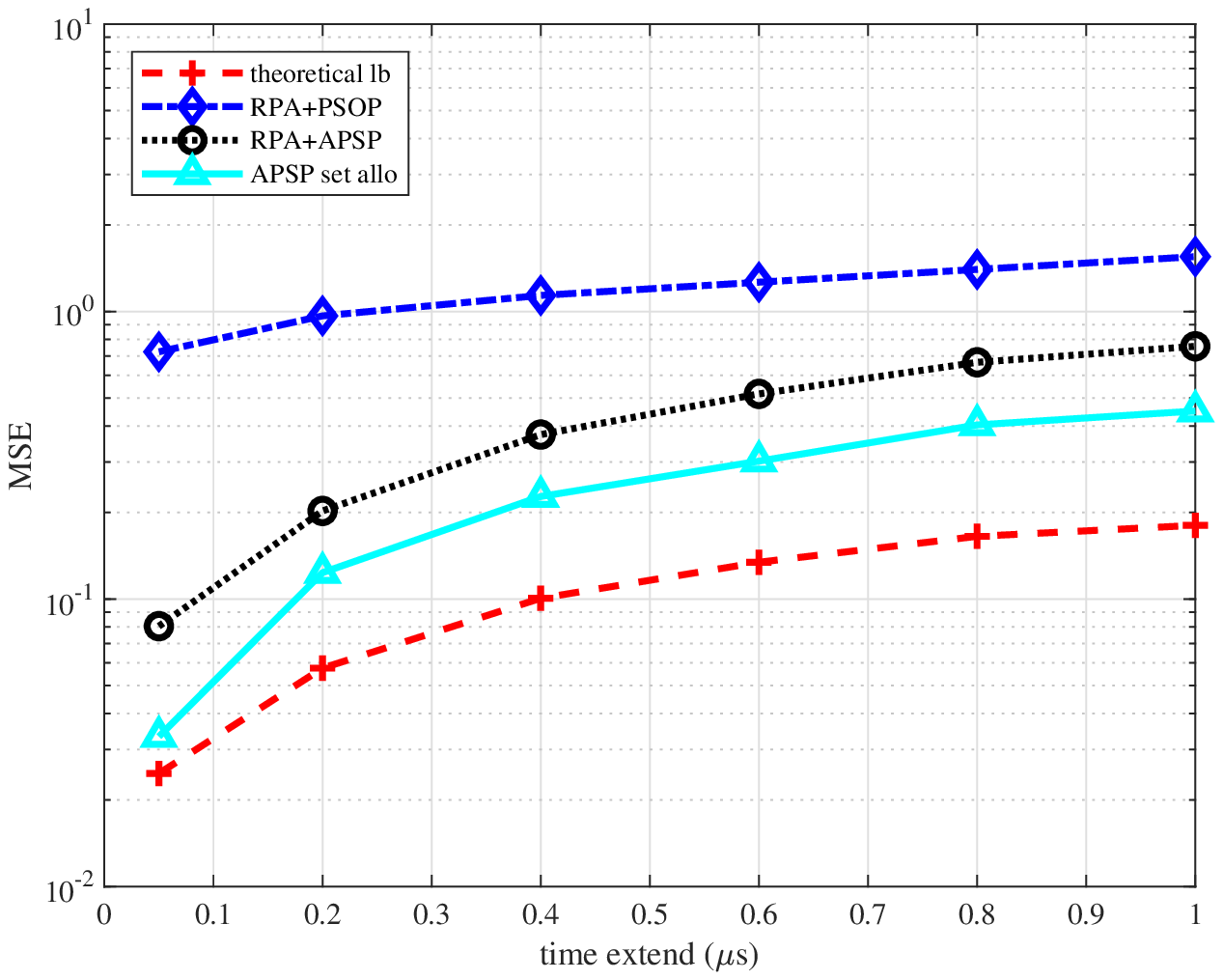}\label{fig:4b}}
\caption{Comparison of MSE-CE of pilot assignment schemes and theoretical lower bound with AP selection versus channel sparsity with $\!L\!=\!10$, $\!K_a\! = \! 50$, $\!\lambda\!=\!0.7$, and $\!Z\!=\!1$. (a) Results are given versus angle spreads with $\zeta\! =\! 0.2~\!\mu \mathrm{s}$. (b) Results are given versus delay spreads with $\varsigma \!=\! 2^\circ$.}
\label{fig:4}
\end{figure*}

  In Figure~\ref{fig:4}, we show the MSE-CE performance of pilot assignment schemes and lower bound versus different channel sparsity.
  In Figure~\ref{fig:4a}, performance comparisons are presented versus different angle spreads. Different delay spreads are considered in Figure~\ref{fig:4b}.
  %The scheme with AP selection significantly outperforms the scheme without AP selection in all angle spread regime and delay spread regime. This is because via AP selection, each UE mainly selects APs closing to it, which are key contributors of channel estimation accuracy.
  The APSP-based RPA scheme outperforms the scheme based on PSOP.
  If the APSP set allocation scheme is adopted, the performance can be further improved and close to its theoretical lower bound especially when the angle spread and delay spread are small.
  Besides, the performance of each scheme improves as angle spread or delay spread decreases, i.e., as channels become more sparse.

\subsubsection{The MSE-CE Performance with Different Numbers of APs and Different AP Selection Coefficients}
\begin{figure*}[!t]
\centering
\subfloat[]{\includegraphics[width=0.49\linewidth]{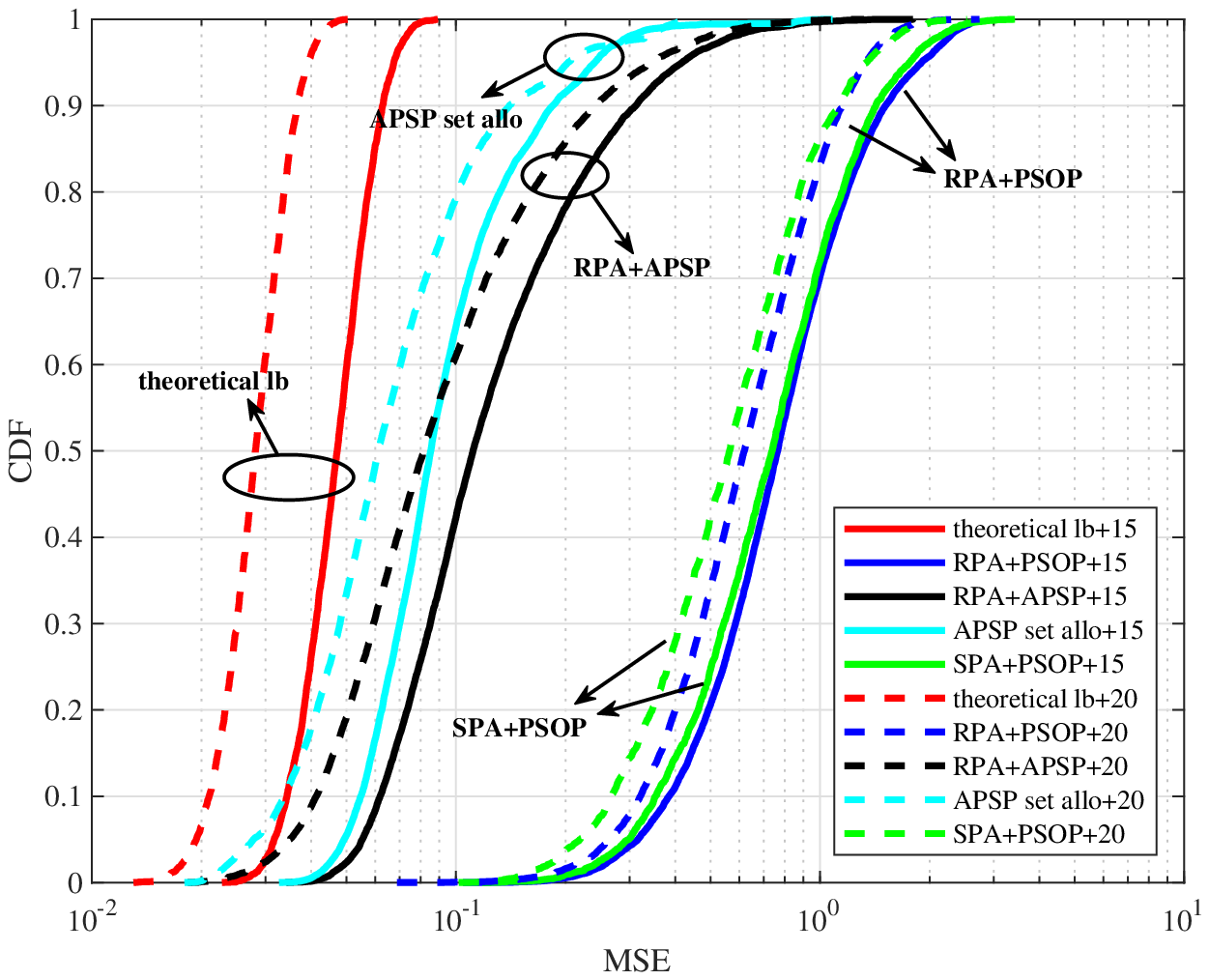}\label{fig:5a}}
\hfil
\subfloat[]{\includegraphics[width=0.49\linewidth]{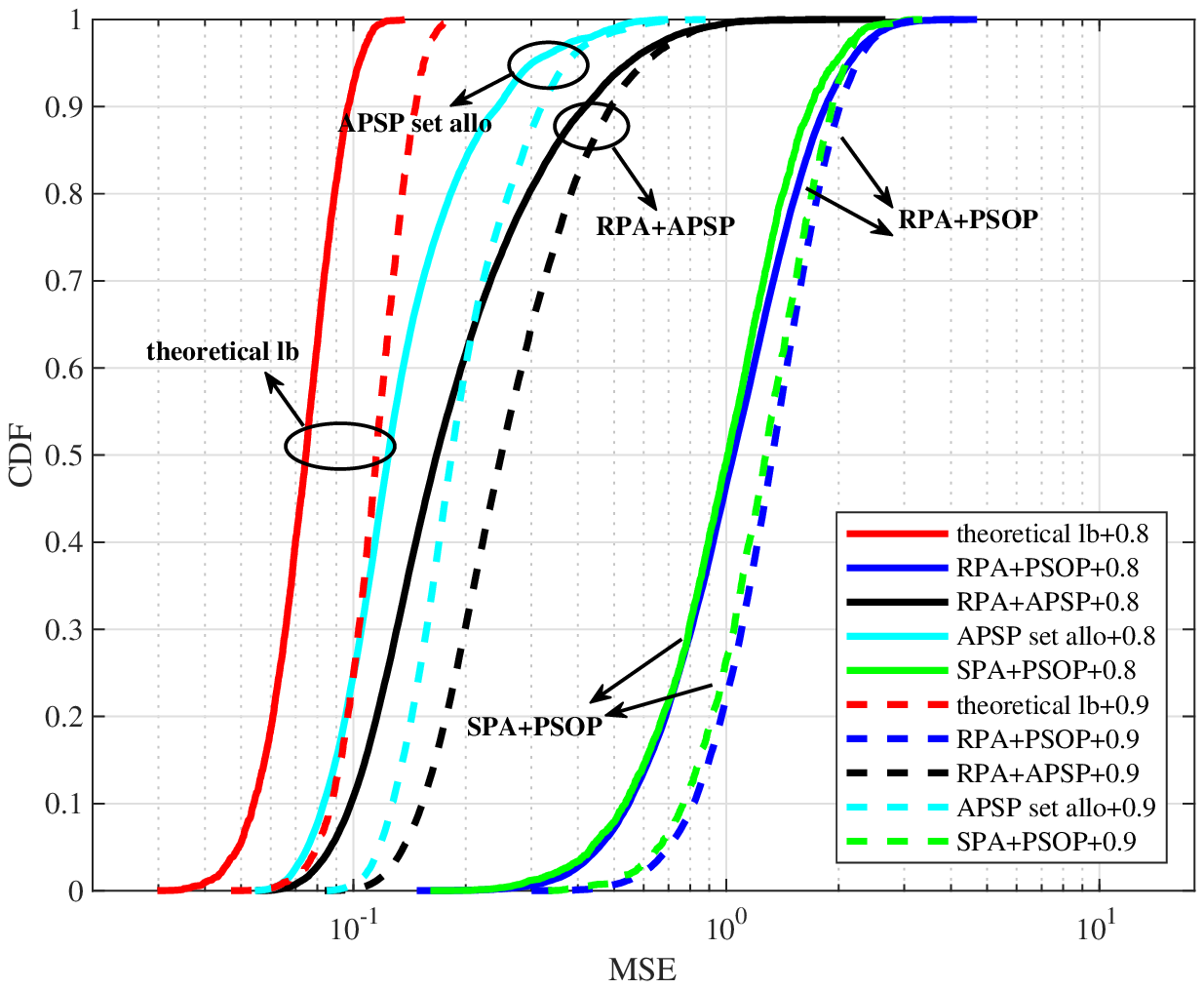}\label{fig:5b}}
\caption{CDFs of MSE-CE for pilot assignment schemes and theoretical lower bounds with AP selection assuming $K_a \!=\! 50$, $\zeta\! =\! 0.2~\!\mu \mathrm{s}$, $\varsigma\! =\! 2^\circ$, and $Z=1$. (a) Results are shown with $\lambda\!=\!0.7$ and $L\!=\!15$ and $20$, respectively. (b) Results are shown with $L\!=\!10$ and $\lambda\!=\!0.8$ and $0.9$, respectively.}
\label{fig:5}
\end{figure*}

  Figure \ref{fig:5} compares the MSE-CE cumulative distributions (CDFs) for pilot assignment schemes and lower bounds.
  In Figure~\ref{fig:5a}, MSE-CE CDFs are presented for different numbers of APs.
  The performance of each scheme improves as the number of APs increases. This is because the macro-diversity provided by many APs reduces the risk that a UE has large distances to all APs.
  Figure~\ref{fig:5b} shows the MSE-CE CDFs of pilot assignment schemes with different AP selection coefficients.
  The performance of each scheme improves as $\lambda$ decreases. This is because APs with better channel estimation performance for a UE are possible to have larger large-scale fading coefficients.
  When $\lambda$ is increased, more APs far from a given UE are taken into account, and the MSE-CE averaged over serving APs of a UE can be worse.
  Besides, compared with the PSOP-based RPA scheme, the SPA scheme~\cite{b15} has limited median gain in crowded scenarios.
  Owing to the reduced pilot interference effect, the performance of the proposed APSP set allocation scheme is close to the lower bound and significantly outperforms other schemes in both median and 95\%-likely performance.

\subsubsection{The SE Performance}
\begin{figure}
  \centering
  \includegraphics[width=0.5\linewidth]{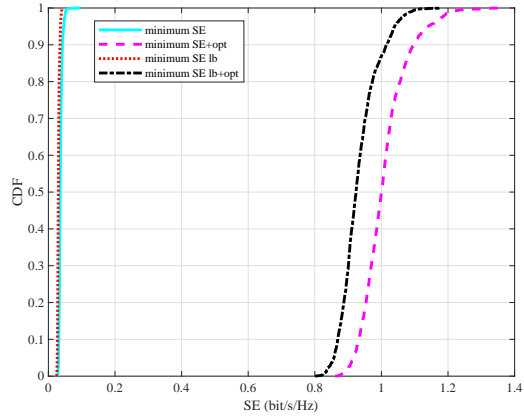}\\
  \caption{CDFs of the minimum SE lower bound among active UEs and corresponding SE with and without power control under the proposed APSP set allocation scheme with $K_a = 50$, $L = 10$, $\lambda = 0.7$, $\zeta = 0.8~\mu \mathrm{s}$, $\varsigma = 8^\circ$, $Z=1$, and 50 taps.}
  \label{fig:7}
\end{figure}

In Figure~\ref{fig:7}, we present the CDFs of the minimum SE and minimum SE lower bound among active UEs with and without power control under the APSP set allocation scheme.
Based on Algorithm \ref{alg:2} with $\epsilon=0.02$, we maximize the minimum SE lower bound among $K_a$ active UEs, which significantly outperforms the case without power control in both median and 95\%-likely performance.
Applying the optimized power coefficients of Problem \eqref{max-min_problem_2} to the real minimum SE among active UEs, it is shown that power control can significantly benefit the minimum SE.

\section{Conclusion} \label{section6}
  In this paper, we considered uplink pilot and data transmission in crowded cell-free massive MIMO-OFDM systems with spatial and frequency correlation.
  For the pilot transmission, we utilized pilot phase shift hopping patterns to identify active UEs. Meanwhile, we derived a closed-form expression of MSE-CE with APSP, and provided an optimal condition of minimizing MSE-CE. According to this condition, we further developed an APSP set allocation scheme to reduce channel estimation error.
  This scheme is insensitive to the active pattern of UEs and does not need to be recomputed frequently.
  Besides, the expressions of SE and a lower bound of SE were derived. For the data transmission, we devised a max-min power control algorithm to maximize the minimum SE lower bound among active UEs. We can obtain the globally optimal solution of this problem by iteratively solving linear programs.
  Significant performance gains in terms of MSE-CE were observed for the proposed APSP set allocation scheme because it can fully utilize the channel sparsity. Compared with the equal power control, our proposed power allocation scheme can improve the minimum SE among active UEs in both median and 95\%-likely performance.
  Hence, the proposed APSP set allocation scheme and the power control scheme are crucial and suitable for crowded cell-free massive MIMO-OFDM systems with frequently and spatially correlated channels.

\begin{appendix}
\section{Proof of Proposition 1} \label{proof_prop_1}
  From the definition of $\mathbf{V}_{N\times N}$ in Proposition 1, we have
    \begin{equation}
      \left[\mathbf{V}_{N\times N}^{H}\mathbf{V}_{N\times N}\right]_{i,j}
      = \frac{1}{N} \sum_{n=0}^{N-1} \exp \left\{-\overline{\jmath}\pi(j-i) \left(\frac{2n}{N}-1 \right) \right\}
      \stackrel{{N \rightarrow \infty}}{=} \delta(j-i),
    \end{equation}
  i.e., $\mathbf{V}_{N\times N}^{H}\mathbf{V}_{N\times N} \stackrel{N \rightarrow \infty}{=} \mathbf{I}_{N\times N}$.
  We have $\mathbf{V}_{M\times M}^{H}\mathbf{V}_{M\times M}
  = \left( \mathbf{I}_{L\times L}^{H} \mathbf{I}_{L\times L} \right) \otimes  \left( \mathbf{V}_{N\times N}^{H} \mathbf{V}_{N\times N} \right)
  \stackrel{N \rightarrow \infty}{=} \mathbf{I}_{M\times M}$.

  From \eqref{g_kls} and \eqref{G_k}, we can obtain
    \begin{equation}\label{vec_G_k}
      \operatorname{vec}\left\{\mathbf{G}_{k}^{\beta}\right \} = \sum_{q=0}^{N_{\rm{cp}}-1}  \int_{-\frac{\pi}{2}}^{\frac{\pi}{2}} \left[\mathbf{f}_{N_{c}, q} \otimes \mathbf{v}_{M}(\theta)\right] \odot \mathbf{a}_{M N_{c}, k}^{\beta}\left(\theta, q T_{s} \right) \mathrm{d} \theta,
    \end{equation}
  where $\mathbf{v}_{M}(\theta)\triangleq\mathbf{1}_{L\times 1}\otimes \mathbf{v}_{N}(\theta)$, $\mathbf{f}_{N_{c}, q}\triangleq\left[1 \; \exp  \left\{ -\overline{\jmath} 2 \pi \frac{1}{N_{c}} q\right\} \!\; \cdots \!\; \exp  \left\{ -\overline{\jmath} 2 \pi \frac{N_{c}-1}{N_{c}} q\right\} \right]^{ T}$,
  $\mathbf{a}_{M N_{c}, k}^{\beta}\left(\theta, q T_{s} \right) \triangleq \mathbf{1}_{N_{c}\times 1} \otimes \mathbf{a}^{\beta}_{L, k}(\theta, q T_{s}) \otimes \mathbf{1}_{N\times 1}$,
  and $\mathbf{a}^{\beta}_{L, k}(\theta, q T_{s})  \triangleq \left[a_{k, 0}^{\beta}(\theta, q T_{s}) \!\;\; a_{k, 1}^{\beta}(\theta, q T_{s}) \cdots  a_{k, L-1}^{\beta}(\theta, q T_{s}) \right]^{T} $.
  The space-frequency domain channel covariance matrix $\mathbf{R}_k^{\beta}$ can be obtained as
    \begin{align} \label{proof_0}
      \mathbf{R}^{\beta}_k
      & = \mathbb{E}\bigg\{ \operatorname{vec}\left\{\mathbf{G}_{k}^{\beta}\right \}  \operatorname{vec}^{H}\left\{\mathbf{G}_{k}^{\beta}\right\}\bigg \}\notag\\
      &= \sum_{q=0}^{N_{\rm{cp}}-1}  \int_{-\frac{\pi}{2}}^{\frac{\pi}{2}} \sum_{q^{\prime}=0}^{N_{\rm{cp}}-1} \int_{-\frac{\pi}{2}}^{\frac{\pi}{2}}
      \left[\mathbf{f}_{N_{c}, q} \otimes \mathbf{v}_{M}(\theta)\right] {\left[\mathbf{f}_{N_{c}, q^{\prime}} \otimes \mathbf{v}_{M}(\theta^{\prime})\right]}^{H}
      \odot \mathbb{E} \left\{ \mathbf{a}_{M  N_{c}, k}^{\beta} \left(\theta, q T_{s} \right) {\left( \mathbf{a}_{M  N_{c}, k}^{\beta} \left(\theta^{\prime} ,q^{\prime} T_{s} \right) \right)}^{ H}  \right\} \mathrm{d} \theta\mathrm{d} \theta^{\prime}.
    \end{align}

  Define $P^{AD}_{k,l}(\theta,q T_s)\triangleq P^{A}_{k, l}(\theta) P^{D}_{k, l}(q T_{s})$. For an arbitrary non-negative integer \emph{d}, let
  $n_{d} \triangleq \lfloor d / M\rfloor$,
  $m_{d}\triangleq  \langle d\rangle_{M}$,
  $r_{m_{d}}\triangleq\lfloor m_{d} / N\rfloor$, and
  $s_{m_{d}}\triangleq\langle m_{d}\rangle_{N}$.
  We have
    \begin{align} \label{proof_1}
      \left[\mathbf{R}^{\beta}_k \right]_{i,j}
      &=\sum_{q=0}^{N_{\rm{cp}}-1} \int_{-\frac{\pi}{2}}^{\frac{\pi}{2}} \sum_{q^{\prime}=0}^{N_{\rm{cp}}-1} \int_{-\frac{\pi}{2}}^{\frac{\pi}{2}}
      \left[\mathbf{f}_{N_{c}, q} \otimes \mathbf{v}_{M}(\theta)\right]_{i} {\left[\mathbf{f}_{N_{c}, q^{\prime}} \otimes \mathbf{v}_{M}(\theta^{\prime})\right]}^{\ast}_{j}
      \mathbb{E} \left\{ \left[\mathbf{a}_{M N_{c}, k}^{\beta}\left(\theta, q T_{s} \right) \right]_{i}      {\left[\mathbf{a}_{M N_{c}, k}^{\beta}\left(\theta^{\prime},q^{\prime} T_{s} \right) \right]}^{\ast}_{j} \right\} \mathrm{d} \theta\mathrm{d} \theta^{\prime} \notag\\
      &\stackrel{(\mathrm a)}{=} \sum_{q=0}^{N_{\rm{cp}}-1} \int_{-\frac{\pi}{2}}^{\frac{\pi}{2}}
      \left[\mathbf{f}_{N_{c}, q} \right]_{n_i}
      \left[\mathbf{f}_{N_{c}, q}\right] ^{\ast}_{n_j}
      \left[\mathbf{v}_{N}(\theta)\right]_{s_{m_{i}}}
      \left[\mathbf{v}_{N}(\theta)\right] ^{\ast} _{s_{m_{j}}}
      \mathbb{E}\left\{
      a_{k, r_{m_i}}^{\beta} \left(\theta , q T_{s}\right)
      \left(a_{k, r_{m_j}}^{\beta}  \left(\theta , q T_{s}\right) \right)^{ \ast}
      \right\} \mathrm{d} \theta \notag\\
      &= \sum_{q=0}^{N_{\rm{cp}}-1} \int_{-\frac{\pi}{2}}^{\frac{\pi}{2}}
      \exp  \left\{-\overline{\jmath} 2 \pi \frac{n_{i}- n_{j}}{N_{\mathrm{c}}} q \right\}
      \exp \left\{- \overline{\jmath} \pi \left(s_{m_i} -s_{m_j} \right) \sin (\theta)\right\}
      \beta\left(k, r_{m_i}\right) P^{AD}_{k,r_{m_i}} (\theta,qT_s)
      \delta \left( r_{m_i} - r_{m_j} \right) \mathrm{d} \theta,
    \end{align}
    where (a) follows from $[\mathbf{A} \otimes  \mathbf{B}]_{i, j}=[\mathbf{A}]_{n_{i}, n_{j}}[\mathbf{B}]_{m_{i}, m_{j}}$ for matrices $\mathbf{A}$ and $\mathbf{B}$. Meanwhile, we have
    \begin{align} \label{proof_1_2}
      &\;\;\;\;\left[ \left(\mathbf{F}_{N_{c} \times N_{\mathrm{cp}}} \otimes \mathbf{V}_{M\times M}\right) \operatorname{diag}\left\{\operatorname{vec} \left\{\boldsymbol{\Upsilon}_{k}^{\beta}\right\}\right\}\left(\mathbf{F}_{N_{c} \times N_{\mathrm{cp}}} \otimes \mathbf{V}_{M\times M}\right)^{H} \right]_{i,j}  \notag\\
      % 1
      &= \sum_{d=0}^{M N_{\mathrm{cp}}-1} \left[ \operatorname{vec} \left\{\boldsymbol{\Upsilon}_{k}^{\beta}\right\} \right]_d
      \left[ \left(\mathbf{F}_{N_{c} \times N_{\mathrm{cp}}} \otimes \mathbf{V}_{M\times M}\right) \right]_{i,d}
      \left[ \left(\mathbf{F}_{N_{c} \times N_{\mathrm{cp}}} \otimes \mathbf{V}_{M\times M}\right) \right]_{j,d}^{\ast}  \notag\\
      % 3
      & =\sum_{n_{d}=0}^{N_{\mathrm{cp}}-1} \sum_{m_{ d}= r_{m_i} }^{r_{m_i} +N-1}
      \left[ \boldsymbol{\Upsilon}_{k,r_{m_d} }^{\beta} \right]_{s_{m_d}, n_{d}}
      \left[\mathbf{F}_{N_{c}\times N_{\mathrm{cp}}} \right]_{ n_i,n_d}
      \left[\mathbf{F}_{N_{c} \times  N_{\mathrm{cp}}} \right]_{ n_j,n_d}^{\ast}
      \left[ \mathbf{V}_{ N \times  N} \right]_{s_{m_i} ,s_{m_d}}
      \left[ \mathbf{V}_{ N \times  N} \right]_{s_{m_j} ,s_{m_d}}^{\ast}
      \delta \left( r_{m_i} -  r_{m_j}\right) \notag\\
      % 4
      & = \sum_{n_{d}=0}^{N_{\mathrm{cp}}-1} \! \sum_{m_{d}=0}^{N-1} \beta\!\left(k, r_{m_i} \right)
      \left(\theta_{m_d+1} \!-\! \theta_{m_d} \right)
      P^{AD}_{k,r_{m_i}}\!(\theta_{m_d},n_{d}T_s)
      \exp  \left\{ -\overline{\jmath} 2 \pi \frac{n_{i}\!-\!n_{j} }{N_{\mathrm{c}}} n_{d} \right\}
      \exp\left\{-\overline{\jmath} \pi  \left(\! s_{m_i} \!\!-\! s_{m_j} \!\right) \sin \left(\theta_{m_d} \right)\right\}
      \delta \!\left(\! r_{m_i}\!\!-\! r_{m_j}\!\right) \notag\\
      % 5
      & \stackrel{{N  \rightarrow \infty}}{=}
      \sum_{q=0}^{N_{\mathrm{cp}}-1} \int_{-\frac{\pi}{2}}^{\frac{\pi}{2}} \beta\left(k, r_{m_i} \right)
      P^{AD}_{k,r_{m_i}}(\theta,q T_s)
      \exp \left\{ - \overline{\jmath} 2 \pi  \frac{n_{i} - n_{j} }{N_{\mathrm{c}}} q \right\}
      \exp  \left\{ - \overline{\jmath} \pi \left( s_{ m_i}  - s_{ m_j} \right)  \sin (\theta) \right\}
      \delta \left( r_{m_i} - r_{m_j} \right) \mathrm{d} {\theta} .
    \end{align}

  Since the power angle-delay spectrum is bounded \cite{b19}, the limit in the first equation of \eqref{proof_1_2} exists. Since \eqref{proof_1} is equal to \eqref{proof_1_2}, $\mathbf{R}_k^{\beta}$ can be obtained as \eqref{relationship_R}.
  The proof of \eqref{relationship_G} is given by
    \begin{align}
      \mathbf{R}^{\beta}_k
      &\stackrel{\stackrel{(a)}{N \rightarrow \infty}}{=}  \left(\mathbf{F}_{N_{c} \times N_{\mathrm{cp}}} \otimes \mathbf{V}_{M\times M}\right)
      \mathbb{E}\left\{ \operatorname{vec} \left\{ \mathbf{H}_{k}^{\beta} \right\}
      \operatorname{vec}^{H} \left\{ \mathbf{H}_{k}^{\beta} \right\}          \right\}
      \left(\mathbf{F}_{N_{c} \times N_{\mathrm{cp}}} \otimes \mathbf{V}_{M\times M}\right)^{H} \notag\\
      &\;\;\stackrel{(b)}{=}
      \mathbb{E}\left\{\operatorname{vec} \left\{ \mathbf{V}_{M\times M} \mathbf{H}_{k}^{\beta}\; \mathbf{F}^{H}_{N_{c} \times N_{\mathrm{cp}}}\right\}
      \operatorname{vec}^{H} \left\{ \mathbf{V}_{M\times M} \mathbf{H}_{k}^{\beta} \;\mathbf{F}^{H}_{N_{c} \times N_{\mathrm{cp}}}\right\} \right\},
    \end{align}
  where (a) follows from \eqref{angle-delay power} and \eqref{relationship_R}, and (b) follows from the fact that $\left( \mathbf{C}^{T}\otimes \mathbf{A}\right) \operatorname{vec}\left\{\mathbf{B} \right\} =\operatorname{vec}\left\{\mathbf{ABC} \right\}$.
  Besides, since $\mathbf{R}^{\beta}_k =\mathbb{E}\left\{ \operatorname{vec} \left\{ \mathbf{G}_{k}^{\beta} \right\} \operatorname{vec}^{H} \left\{ \mathbf{G}_{k}^{\beta}\right\} \right\}$, we can obtain \eqref{relationship_G}.

\section{Proof of Proposition 2} \label{proof_prop_2}
  Considering the non-negative property of the angle-delay domain channel power distribution, it is satisfied that in \eqref{a_MSE_CE} the term
  $R_{\mathcal{K}_a^{i,j},p,m,q} = \frac{{{\left[ \mathbf{\Upsilon }_{\mathcal{K}_{a}^{i,j}}^{\beta,0} \right]}_{m,q}}}
  {{{\left[ \mathbf{\Upsilon }_{\mathcal{K}_{a}^{i,j}}^{\beta} \right]}_{m,q}}
  +\sum\limits_{\stackrel{j'=0}{j'\neq j}  }^{K_a-1}
  \delta\left(\left\langle\mathcal{P}_{i,j'}^{p}\right\rangle_{Z}
  -\left\langle \mathcal{P}_{i,j}^{p} \right\rangle_{ Z}\right)
  {{{\left[ \mathbf{\Upsilon }_{\mathcal{K}_{a}^{i,{j}^{\prime}}}^{\beta,\left\lfloor {\mathcal{P}_{i,j'}^{p}}/{Z} \right\rfloor - \left\lfloor
  {\mathcal{P}_{i,j}^{p}}/{Z} \right\rfloor}  \right]}_{ m,q}}
  }  + \frac{1}{{{\rho }_{p}}Z}}  \in  [0,1)$.
  Then, we have $\varepsilon_{\mathcal{K}_a^{i,j},p,m,q}^{0} = {{\left[ \mathbf{\Upsilon }^{0}_{\mathcal{K}_{a}^{i,j}}\right]}_{m,q}}\left(1- R_{\mathcal{K}_a^{i,j},p,m,q} \right) \geq 0$.
  When ${{\left[ \mathbf{\Upsilon }^{0}_{\mathcal{K}_{a}^{i,j}}\right]}_{m,q}}=0$, $\varepsilon_{\mathcal{K}_a^{i,j},p,m,q}^{0}$ can achieve the minimum value $0$;
  when ${{\left[ \mathbf{\Upsilon }^{0}_{\mathcal{K}_{a}^{i,j}}\right]}_{m,q}}>0$, $\varepsilon_{\mathcal{K}_a^{i,j}\!,p,m,q}^{0}$ can be minimized if and only if
  $R_{\mathcal{K}_a^{i,j}\!,p,m,q}$ is maximized,
  i.e., $\sum\limits_{\stackrel{j'=0}{j'\neq j}  }^{K_a-1}
  \!\delta\!\left(\left\langle\mathcal{P}_{i,j'}^{p} \!\right\rangle_{\! Z}
  \!\!-\!\left\langle \mathcal{P}_{i,j}^{p} \right\rangle_{\! Z}\right)
  {{{\!\left[ \!\mathbf{\Upsilon }_{\mathcal{K}_{a}^{i,{j}^{\prime}}}^{\beta,\left\lfloor {\mathcal{P}_{i,j'}^{p}}\!/{Z} \right\rfloor - \left\lfloor
  {\mathcal{P}_{i,j}^{p}}/{Z} \right\rfloor}  \right]}_{m,q}}
  } \!= 0$.
  Hence, we can conclude the minimum condition of $\varepsilon_{\mathcal{K}_a^{i,j} ,p,m,q}^{0}$ as
  ${{\left[ \mathbf{\Upsilon }^{0}_{\mathcal{K}_{a}^{i,j}}\right]}_{m,q}} {{\left[ \mathbf{\Upsilon }^{0}_{\mathcal{K}_{a}^{i,j}}\right]}_{m,q}}
  {{{\left[ \mathbf{\Upsilon }_{\mathcal{K}_{a}^{i,{j}^{\prime}}}^{\beta,\left\lfloor {\mathcal{P}_{i,j'}^{p}}/{Z} \right\rfloor - \left\lfloor
  {\mathcal{P}_{i,j}^{p}}/{Z} \right\rfloor} \right]}_{ m,q}}
  }=0$ for $  \left\langle\mathcal{P}_{i,j'}^{p} \right\rangle_{Z}
  =\left\langle \mathcal{P}_{i,j}^{p}  \right\rangle_{Z}$, $j'=0,1,\cdots,K_a-1$, and $j'\neq j$.

  For any choice of $i,j,p,m, \text{and}\; q$, $\varepsilon_{\mathcal{K}_a^{i,j} ,p,m,q}^{0}$ should be minimized. The optimal condition can be expressed as $\mathbf{\Upsilon }^{0}_{k} \odot  \mathbf{\Upsilon }^{0}_{k}  \odot \mathbf{\Upsilon }_{{{k}^{\prime}}}^{\beta ,\left\lfloor{{\phi }_{{{k}^{\prime}}}}/Z\right\rfloor-\left\lfloor{{\phi }_{{{k}}}}/Z\right\rfloor} = \mathbf{0}$ for $\left\langle {{\phi }_{{{k}^{\prime}}}}\right\rangle_Z = \left\langle {{\phi }_{{{k}}}}\right\rangle_Z$, $k^{\prime}\in \mathcal{K}$, and $k\neq k^{\prime}$.
  It is a necessary and sufficient condition, but it can be expressed as other forms.
  The reason why we choose $\mathbf{\Upsilon }_{k}^{0}\odot \mathbf{\Upsilon }_{k}^{0}\odot\mathbf{\Upsilon }_{{{k}^{\prime}}}^{\beta ,\left\lfloor{{\phi }_{{{k}^{\prime}}}}/Z\right\rfloor-\left\lfloor{{\phi }_{{{k}}}}/Z\right\rfloor }$
  but not $\mathbf{\Upsilon }_{k}^{0}\odot\mathbf{\Upsilon }_{{{k}^{\prime}}}^{\beta ,\left\lfloor{{\phi }_{{{k}^{\prime}}}}/Z\right\rfloor-\left\lfloor{{\phi }_{{{k}}}}/Z\right\rfloor }$ is because we want to make the interference $\mathbf{\Upsilon }_{{{k}^{\prime}}}^{\beta ,\left\lfloor{{\phi }_{{{k}^{\prime}}}}/Z\right\rfloor-\left\lfloor{{\phi }_{{{k}}}}/Z\right\rfloor }$ smaller when $\mathbf{\Upsilon }_{k}^{0}$ is large.
  In this way, ${{{\overline{\varepsilon }}}^{0}}$ can be smaller.

  Based on \eqref{a_MSE_CE} and \eqref{a_MSE_CE2}, we can obtain
    \begin{align}\label{proof_2}
      {{\mathbb{E}}_{\mathcal{U},{{\mathcal{K}}_{{a}}} ,\mathcal{P}}} (\varepsilon^{0} )
      &  = \sum\limits_{i=0}^{{{N}_{{ {\mathcal{K}}_{a}}}- 1}}  {\sum\limits_{j=0}^{{{K}_{a}-1}}
      {\sum\limits_{p=0}^{{{N}_{\phi } -1}}  {\frac{1}
      {{{N}_{{{\mathcal{K}}_{ a}}}} {{K}_{ a}}{{N}_{\phi }} {{N}_{c}} {\left|\mathcal{B}_{ {\mathcal{K}}^{i,j}_{a}}\right|} }}
      {\sum\limits_{m=0}^{M -1}    {\sum\limits_{q=0}^{{{N}_{\mathrm{cp}}} -1}
      {\left\{ \varepsilon_{\mathcal{K}_a^{i,j},p,m,q}^{0} \right\}}}} }}  \notag \\
      & {\geq}   \sum\limits_{i=0}^{{{N}_{{{\mathcal{K}}_{a}}} -1}}  {\sum\limits_{j=0}^{{{K}_{a} -1}}
      {\sum\limits_{p=0}^{{{N}_{\phi } -1}} {\frac{1}
      {{{N}_{{{\mathcal{K}}_{a}}}}{{K}_{a}}{{N}_{\phi }} {{N}_{c}} {\left|\mathcal{B}_{ {\mathcal{K}}^{i,j}_{a}}\right|} }}
      {\sum\limits_{m=0}^{M-1}
      {\sum\limits_{q=0}^{{{N}_{\mathrm{cp}}}-1}
      {\left\{  {{\left[ \mathbf{\Upsilon }^{0}_{\mathcal{K}_{a}^{i,j}}\right]}_{m,q}} -
      \frac{{{\left[ \mathbf{\Upsilon }^{0}_{\mathcal{K}_{a}^{i,j}}\right]}_{m,q}}
      {{\left[ \mathbf{\Upsilon }_{\mathcal{K}_{a}^{i,j}}^{\beta,0} \right]}_{m,q}}}
      {{{\left[ \mathbf{\Upsilon }_{\mathcal{K}_{a}^{i,j}}^{\beta} \right]}_{m,q}}  + \frac{1}{{{\rho }_{p}}Z}} \right\}}}}
      }}   \notag \\
      & \stackrel{(\mathrm a)}{=} {\sum\limits_{k=0}^{{K-1}} {\frac{1}
      {K {{N}_{c}} {\left|\mathcal{B}_{k}\right|} }}
      {\sum\limits_{m=0}^{M-1}
      {\sum\limits_{q=0}^{{{N}_{\mathrm{cp}}}-1}
      {\left\{ {{\left[ \mathbf{\Upsilon }^{0}_{k}\right]}_{m,q}}-
      \frac{{{\left[ \mathbf{\Upsilon }^{0}_{k}\right]}_{m,q}}
      {{\left[ \mathbf{\Upsilon }_{k}^{\beta,0} \right]}_{m,q}}}
      {{{\left[ \mathbf{\Upsilon }_{k}^{\beta} \right]}_{m,q}} + \frac{1}{{{\rho }_{p}Z}}} \right\}}}}
      }  \notag \\
      &=\left[{{{\mathbb{E}}_{\mathcal{U},{{\mathcal{K}}_{a}},\mathcal{P}}} (\varepsilon^{0} )} \right]_{\min },
    \end{align}
  where (a) follows from the fact that when $\varepsilon_{\mathcal{K}_a^{i,j},p,m,q}^{0}$ is minimized, i.e., when the effect of pilot interference is eliminated, the average operation accounting for all possible active patterns and all types of phase shift selection is the same as the average operation over $K$ UEs in the network \cite{b9}.
  Then, we have $\overline\varepsilon^{0}_{\min}= \sum\limits_{K_{a} =1}^{K} { p(K_a|K)  \left[{{{\mathbb{E}}_{\mathcal{U},{{\mathcal{K}}_{a}},\mathcal{P}}} (\varepsilon^{0} )} \right]_{\min } }$.

\section{Derivation of \eqref{SE_lb}} \label{proof_prop_3}
  We can rewrite \eqref{r_ks} as
  \begin{equation}
      {{r}_{k,s}}=\sum\limits_{l=0}^{L-1}{\nu_{k,l}{{r}_{k,l,s}}}
      = \sqrt{{{\rho }_{u}}}\sqrt{{{\eta }_{k}}} \;\mathbb{E}\left\{ \left(\mathbf{c}_{k,s}^{\beta,0 }\right)^{H} {\mathbf{g}}_{k,s}^{\beta } \right\}{{x}_{k,s}}
      + \text{Inf}_{\text{sum}}^{\; '} ,
   \end{equation}
  where the interference term is given by
    \begin{equation}
      \text{Inf}_{\text{sum}}^{\; '} =  \sqrt{{{\rho }_{u}}{{\eta }_{k}}} \left( \left( \mathbf{c}_{k,s}^{\beta,0 } \right)^{H}  {\mathbf{g}}_{k,s}^{\beta }  - \mathbb{E} \left\{  \left(\mathbf{c}_{k,s}^{\beta,0 } \right)^{H} {\mathbf{g}}_{k,s}^{\beta }  \right\}  \right) {{x}_{k,s}}
      +\sqrt{{{\rho }_{u}}} \sum\limits_{{k}'\in \;\mathcal{U}_{\mathcal{K}_{a}}^i \backslash \left\{ k \right\} } {\sqrt{{{\eta }_{{{k}'}}}}\left( \mathbf{c}_{k,s}^{\beta,0 }\right)^{ H}  \mathbf{g}_{{k}',s}^{\beta }{{x}_{{k}',s}}}
      +\left(\mathbf{c}_{k,s}^{\beta,0 }\right)^{H} {{\mathbf{w}}_{s}}.
    \end{equation}
  We have $\mathbb{E}\left\{\text{Inf}_{\text{sum}}^{\; '}\right\}=0$. Since the transmitted signal of UE \emph{k} is independent of the signals of other UEs and receiver noise, the interference is uncorrelated with the transmitted signal, i.e.,
    \begin{equation}
      \mathbb{E}\left\{ x_{k,s}^{\ast} \text{Inf}_{\text{sum}}^{\; '} \right\} = \sqrt{{{\rho }_{u}}} \sqrt{{{\eta }_{k}}}\;
      \mathbb{E}\left\{  \left( \mathbf{c}_{k,s}^{\beta,0 } \right)^{ H}   {\mathbf{g}}_{k,s}^{\beta } - \mathbb{E} \left\{  \left( \mathbf{c}_{k,s}^{\beta,0 } \right)^{H}  {\mathbf{g}}_{k,s}^{\beta }  \right\} \right\}
      \mathbb{E}\left\{ |{{x}_{k,s}} |^{2} \right\} = 0.
    \end{equation}

  The variance of the interference term is represented as
    \begin{equation}
      \mathbb{E}\left\{ {{\left| \text{Inf}_{\text{sum}}^{\; '}  \right|}^{2}} \right\}
      ={{\rho }_{u}} \sum\limits_{{k}'\in \mathcal{U}^{i}_{{\mathcal{K}}_{a}} }
       {{{\eta }_{{{k}'}}}
      \mathbb{E} \left\{ {{\left| \left(\mathbf{c}_{k,s}^{\beta,0 }\right)^{ H} \mathbf{g}_{{k}',s}^{\beta } \right|}^{2}}\right\}}
      - {{\rho }_{u}}{{\eta }_{k}}
       \left|\mathbb{E} \left\{  {{ \left(\mathbf{c}_{k,s}^{\beta,0 }\right)^{ H}  {\mathbf{g}}_{k,s}^{\beta }}}\right\} \right|^{2}
      +\mathbb{E} \left\{ \left\| \mathbf{c}_{k,s}^{\beta,0}\right\|^{2}\right\} .
    \end{equation}
  It follows from the independence between each of the zero-mean transmitted signals and the independence between signals and channels. Taking the OFDM CP overhead and pilot overhead into account, according to Corollary 1.3 in \cite{b26}, we can obtain the SE lower bound as shown in \eqref{SE_lb}.
\end{appendix}

\Acknowledgements{The work of Y. Wu was supported in part by the National Key R\&D
Program of China under Grant 2018YFB1801102,  JiangXi Key R\&D Program
under Grant 20181ACE50028, National Science Foundation
(NSFC) under Grant 61701301, the open research project of State Key Laboratory of Integrated
Services Networks (Xidian University) under Grant ISN20-03, and Shanghai Key Laboratory of Digital Media Processing and Transmission (STCSM18DZ2270700).
The work of Y. Wang was supported by the National Natural Science Foundation of China (No.61931019).
The work of W. Zhang was supported by Shanghai Key Laboratory of Digital Media Processing and Transmission (STCSM18DZ2270700).}

\end{document}